\documentclass[aps,prb,secnumarabic,nobibnotes,twocolumn,reprint,superscriptaddress]{revtex4-2}
\usepackage{dsfont}
\usepackage{fontenc}
\usepackage{verbatim}
\usepackage{amsfonts}
\usepackage{mathrsfs}
\usepackage{amsmath}
\usepackage{color}
\usepackage{natbib}
\usepackage{graphicx}
\usepackage{bm}
\usepackage{amssymb}
\usepackage{xspace}
\usepackage{epstopdf}
\usepackage{dcolumn}
\usepackage{multirow}
\usepackage{tabularx}
\usepackage{bbding}
\usepackage{booktabs}
\usepackage[colorlinks=true, letterpaper=true, pdfstartview=FitV, linkcolor=blue, citecolor=blue, urlcolor=blue]{hyperref}

\begin{document}

\title{Manipulating anomalous transport via crystal symmetry in two-dimensional altermagnets}

\author{Dan Li}
\thanks{These authors contributed equally to this work.}
\address{Research Center for Quantum Physics and Technologies, School of Physical Science and Technology, Inner Mongolia University, Hohhot 010021, China}
\address{Key Laboratory of Semiconductor Photovoltaic Technology and Energy Materials at Universities of Inner Mongolia Autonomous Region, Inner Mongolia University, Hohhot 010021, China}

\author{Shuaiyu Wang}
\thanks{These authors contributed equally to this work.}
\address{Research Center for Quantum Physics and Technologies, School of Physical Science and Technology, Inner Mongolia University, Hohhot 010021, China}

\author{Jiabing Chen}
\address{Research Center for Quantum Physics and Technologies, School of Physical Science and Technology, Inner Mongolia University, Hohhot 010021, China}
\address{Key Laboratory of Semiconductor Photovoltaic Technology and Energy Materials at Universities of Inner Mongolia Autonomous Region, Inner Mongolia University, Hohhot 010021, China}

\author{Zeling Li}
\address{Research Center for Quantum Physics and Technologies, School of Physical Science and Technology, Inner Mongolia University, Hohhot 010021, China}
\address{Key Laboratory of Semiconductor Photovoltaic Technology and Energy Materials at Universities of Inner Mongolia Autonomous Region, Inner Mongolia University, Hohhot 010021, China}

\author{Chaoxi Cui}
\email{cuichaoxi@gmail.com}
\affiliation{Key Lab of Advanced Optoelectronic Quantum Architecture and Measurement (MOE), Beijing Key Lab of Nanophotonics \& Ultrafine Optoelectronic Systems, and School of Physics, Beijing Institute of Technology, Beijing 100081, China}

\author{Lei Li}
\address{Research Center for Quantum Physics and Technologies, School of Physical Science and Technology, Inner Mongolia University, Hohhot 010021, China}

\author{Lei Wang}
\address{Research Center for Quantum Physics and Technologies, School of Physical Science and Technology, Inner Mongolia University, Hohhot 010021, China}
\address{Inner Mongolia Key Laboratory of Microscale Physics and Atom Innovation, Inner Mongolia University, Hohhot 010021, China}

\author{Zhi-Ming Yu}
\affiliation{Key Lab of Advanced Optoelectronic Quantum Architecture and Measurement (MOE), Beijing Key Lab of Nanophotonics \& Ultrafine Optoelectronic Systems, and School of Physics, Beijing Institute of Technology, Beijing 100081, China}

\author{Xiaodong Zhou}
\email{zhouxiaodong@tiangong.edu.cn}
\address{School of Physical Science and Technology, Tiangong University, Tianjin 300387, China}

\author{Xiao-Ping Li}
\email{xpli@imu.edu.cn}
\address{Research Center for Quantum Physics and Technologies, School of Physical Science and Technology, Inner Mongolia University, Hohhot 010021, China}
\address{Key Laboratory of Semiconductor Photovoltaic Technology and Energy Materials at Universities of Inner Mongolia Autonomous Region, Inner Mongolia University, Hohhot 010021, China}

\begin{abstract}

Anomalous transports, including the anomalous Hall effect (AHE) and anomalous Nernst effect (ANE), are typical manifestations of time-reversal-symmetry-breaking responses in materials. In general, the two Hall states with opposite Hall conductivities can be regarded as time-reversal pairs coupled to magnetic order, and switching between them relies on reversing the magnetization via an external magnetic field or electric current. Here, we introduce an approach for manipulating anomalous transport through crystal symmetry engineering in two-dimensional (2D) altermagnetic systems. Through symmetry analysis and lattice models, we show that for systems with out-of-plane Néel vectors, breaking the symmetry that connects two magnetic sublattices will induce an anomalous Hall response. Crucially, the signs of the anomalous Hall and anomalous Nernst conductivities can be flexibly controlled via symmetry-breaking perturbations, enabling the realizing tunable sign-reversible anomalous transport. 
Using first-principles calculations, we investigate crystal-symmetry-controlled anomalous transport in the 2D altermagnetic materials V$_{2}$STeO and Cr$_{2}$O$_{2}$.  
The results indicate that these systems with out-of-plane Néel vectors exhibit pronounced AHE and ANE signals under uniaxial strain. Remarkably, the sign of the anomalous transport coefficients can be reversed by simply rotating the strain direction by $\pi/2$, 
thus enabling the switching of anomalous transverse currents without an external magnetic field. Our findings provide a new strategy to manipulate anomalous transport, and should have significant potential applications. 

\end{abstract}

\maketitle

\section{Introduction}\label{intro}

The intrinsic anomalous transport phenomena originating from band geometric quantities (Berry curvature) in magnetic materials, including the anomalous Hall effect (AHE) and its thermoelectric analogue, the anomalous Nernst effect (ANE), have always been a long-standing topic in condensed matter physics~\cite{RevModPhys.82.1539, vsmejkal2022anomalous, PhysRevLett.97.026603}. Upon integrating the Berry curvature of Bloch bands over a closed manifold, a quantized topological invariant (Chern number) emerges, giving rise to the quantum anomalous Hall effect~\cite{ezawa2013quantum, RevModPhys.95.011002}, which exhibits quantized anomalous transport signals, subsequently trigger widespread interest in the topological classification of matter~\cite{RevModPhys.88.035005, RevModPhys.82.3045, RevModPhys.83.1057, RevModPhys.88.021004, PhysRevB.78.195125, PhysRevLett.106.106802, benalcazar2017quantized, PhysRevLett.119.246402, RevModPhys.90.015001}. The anomalous transport phenomena were first studied
in ferromagnetic materials~\cite{PhysRevLett.88.207208, fang2003anomalous, PhysRevLett.92.037204, PhysRevLett.97.126602, PhysRevB.77.165103, PhysRevLett.100.016601, PhysRevB.81.054414, PhysRevLett.93.226601, PhysRevLett.119.056601, PhysRevB.90.054422, sakai2018giant} and in turn attracted attention in antiferromagnetic systems~\cite{vsmejkal2022anomalous, PhysRevLett.112.017205, nakatsuji2015large, PhysRevApplied.5.064009, nayak2016large, higo2018anomalous, PhysRevB.101.094404, feng2022anomalous, ghimire2018large, machida2010time, suzuki2016large, takahashi2018anomalous,Ikhlas2017,XK-Li2017,XD-Zhou2019,XD-Zhou2020,XD-Zhou2023}. Among these investigations, a promising direction lies in manipulating the anomalous Hall conductivity to achieve sign reversal, which  provides a foundation for the design of AHE-based spintronic devices~\cite{kim2022ferrimagnetic, RevModPhys.90.015005}. 

Generally, the anomalous Hall conductivity is expressed as an antisymmetric second-rank tensor $\sigma_{ij}^{A}$, which can be characterized by an axial Hall vector $\mathbf{h}=\left(\sigma_{zy}^{A},\sigma_{xz}^{A},\sigma_{yx}^{A}\right)$~\cite{vsmejkal2022anomalous}. Crucially, the components of $\mathbf{h}$ change sign under the time-reversal transformation $\mathcal{T}$, thereby enabling a sign reversal of the Hall conductivity. Experimentally, this is realized by flipping the magnetization (or Néel vector) with external perturbations such as a magnetic field or other spin-polarized current~\cite{li2020giant, PhysRevB.101.140405, zhu2024unveiling}, an action that performs the same operation as time reversal and thus changes the sign of anomalous transport. 
However, reversing the magnetic order of a crystal by means of an external field requires overcoming a sizable magnetic-anisotropy energy (MAE) barrier~\cite{hirjibehedin2007large, natterer2017reading} and is inevitably impeded by stray fields, Joule heat and quantum tunnelling~\cite{kim2022ferrimagnetic, thomas1996macroscopic, miao2011tunneling}. Consequently, the search for new routes to manipulate anomalous transport has become one of the key objectives in spintronics.

Recently, a new magnetic phase known as altermagnet (AM) has been proposed~\cite{PhysRevX.12.040501, PhysRevX.12.031042, vsmejkal2020crystal, mazin2021prediction,MaHY2021}, which characterized by non-relativistic alternating spin splitting in momentum space and collinear compensated magnetic orders in real space. 
AM combines the advantages of ferromagnetic and antiferromagnetic orders, 
making them particularly suitable for spintronic applications~\cite{bai2024altermagnetism, song2025altermagnets, jungwirth2024altermagnets, guo2025spin}, including time-reversal-symmetry-breaking anomalous transport~\cite{han2024electrical, PhysRevLett.130.036702, feng2022anomalous, wang2023emergent,Smejkal2020,XD-Zhou2024,Takagi_FeS2_NM2025,Jeong2025, PhysRevB.109.224430, reichlova2024observation, badura2025observation}. Importantly, altermagnetic order is essentially a magnetic crystal order determined by the magnetic-order (Néel) vector and crystal symmetry~\cite{PhysRevX.12.040501, PhysRevX.15.021083, zhou2025manipulation, zhou2024crystal}, in which the two magnetic sublattices are connected through rotation or mirror symmetry operations. This indicates that, in addition to tuning the Néel order as in conventional antiferromagnets, AM can modulate magnetic symmetry and the corresponding anomalous transport by breaking crystalline symmetry. Such a magnetic modulation strategy may offer unprecedented opportunities for crystal-level anomalous-transport engineering.

In this work, we propose a strategy to realize the anomalous transport and its sign-reversal in two-dimensional (2D) altermagnets via breaking the crystal symmetry.  
We show that a 2D AM system with out-of-plane magnetization can exhibit an intrinsic anomalous Hall (Nernst) effect once the symmetry between the two magnetic sublattices is broken. Moreover, two distinct symmetry-breaking perturbations related by symmetry can induce opposite anomalous transport, thereby enabling sign-switchable control of the anomalous Hall and Nernst conductivities in a fully predictable manner. A tight-binding (TB) model is constructed to demonstrate the feasibility of our proposal. 
To realize this mechanism, we further identify two 2D altermagnetic materials: monolayer V$_{2}$STeO and Cr$_{2}$O$_{2}$. In pristine states with out-of-plane magnetization, both systems exhibit vanishing anomalous transport owing to the strict constraint of the $C_{4z} \mathcal{T}$ symmetry connecting opposite-spin sublattices.
However, once the $C_{4z}$ rotation symmetry is broken via uniaxial strain, both materials can produce a finite anomalous Hall response. Remarkably, uniaxial strain applied along the $a$ or $b$ axis generates anomalous Hall and anomalous Nernst conductances that are equal in magnitude but opposite in sign, enabling full sign-reversal control of the transverse transport. 
Finally, the design of a cascaded Hall structure based on symmetry-controlled transport mechanisms is discussed. Our work establishes a promising route to manipulating anomalous transport via crystal symmetry, laying the groundwork for its experimental realization and for spintronic applications.
\begin{figure}[t]
\includegraphics[width=8.4cm]{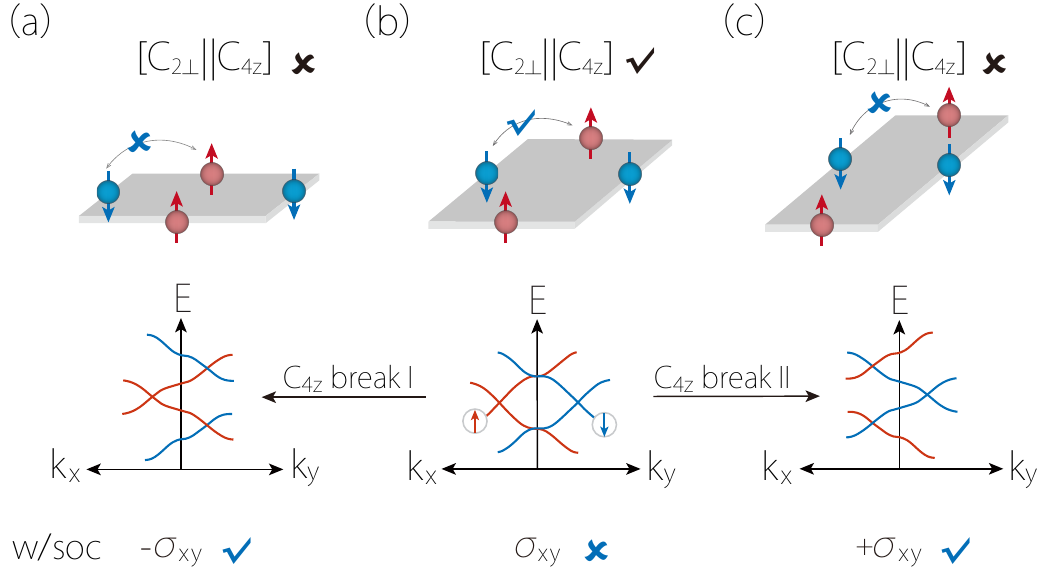}
\caption{Schematic of crystal-symmetry-controlled anomalous transport in a $\left[C_{2\perp}||C_{4z}\right]$ altermagnetic example. (b) illustrates the pristine altermagnet and its band characteristics. (a) and (c) show the possible band evolution and corresponding anomalous Hall responses after introducing two distinct symmetry-breaking perturbations related by $C_{4z}\mathcal{T}$, respectively.
\label{fig1}}
\end{figure}

\section{Rationale and lattice model}\label{analyze}

To achieve manipulation of anomalous transport in 2D AM system via crystal symmetry, the following two conditions must be satisfied: (i) The pristine system exhibits no anomalous Hall response; (ii) Upon breaking the crystal symmetry connecting the two spin sublattices, an anomalous Hall signal emerges. It can be shown that conditions above are directly satisfied once the Néel vector of a 2D AM system points in the out-of-plane direction. As pointed out by Sheoran \textit{et al.}~\cite{PhysRevB.111.184407}, the AHE is always forbidden in 2D AMs when the Néel vector points out of plane, thereby satisfying our Condition (i). This can be easily understood through a brief argument based on the reduction of the spin group to the magnetic group.
For a 2D AM, owing to the 2D geometry and the in-plane nature of the crystal momentum $k$, the opposite-spin sublattices can be connected only by in-plane twofold rotations $C_{2\parallel}$ (e.g., $C_{2x}$, $C_{2y}$, and $C_{2xy}$), vertical mirror operations $M_{\parallel}$ (e.g., $M_x$, $M_y$, and $M_{xy}$), and out-of-plane symmetries $\{C_{nz},\, S_{nz}\}$ with $n=3,4,6$ \cite{BailingTypeIV}. As a result, the general symmetry operations that connect the two spin sublattices in the spin space group can be written as

\begin{eqnarray}\label{ssg11}
\bigl\{\left[C_{2\perp}||C_{2\parallel}\right],\left[C_{2\perp}||C_{nz}\right],\left[C_{2\perp}||M_{\parallel}\right], \left[C_{2\perp}||S_{nz}\right]\bigl\}
\end{eqnarray}
Here, $C_{2\perp}$ denotes a twofold rotation in spin space about an axis perpendicular to the collinear spins. Upon introducing spin-orbit coupling (SOC) and setting the Néel vector out of plane [see Supplemental Material (SM)~\cite{sm} for details], the spin group operations in Set (\ref{ssg11}) reduce to the following magnetic group operations:
\begin{eqnarray}\label{msg3}
\left\{ C_{2\parallel},C_{nz}\mathcal{T},M_{\parallel},S_{nz}\mathcal{T}\right\}. 
\end{eqnarray}
Here, $\mathcal{T}$ is the time-reversal symmetry.
Notably, any of the magnetic group operations in Set (\ref{msg3}) forces the anomalous Hall conductivity to vanish, consistent with the conclusions of Sheoran \textit{et al.}~\cite{PhysRevB.111.184407}.

When the symmetry connecting the magnetic sublattices is broken [i.e., Condition (ii) is satisfied], the magnetic group operations that forbid the AHE are no longer preserved, leading to the emergence of the AHE. This result is straightforward to understand, as the system essentially transitions from an AM to a ferrimagnet, thereby exhibiting an anomalous Hall response akin to that of a ferromagnet with out-of-plane magnetization. Since the out-of-plane magnetization and the anomalous Hall conductivity tensor share identical symmetry transformation properties in 2D ferromagnets, systems permit that a non-zero out-of-plane magnetization inherently allows for the anomalous Hall effect~\cite{RevModPhys.82.1539}. Based on this result, anomalous transport can be readily induced by applying symmetry-breaking perturbations.

Next, we show that the sign of the induced anomalous Hall response can be fully reversed by switching the external perturbation. Because a symmetry operation $g\in\left\{ C_{2\parallel},C_{nz}\mathcal{T},M_{\parallel},S_{nz}\mathcal{T}\right\} $ connecting the magnetic sublattices  enforces a vanishing anomalous Hall conductivity $\sigma=0$ in the pristine system. Introducing a perturbation $\varepsilon$  breaking $g$ induces a finite response, $\sigma(\varepsilon)\neq0$. Importantly, the system can host two energetically equivalent symmetry-broken states connected by $g$: one realized by the perturbation $\varepsilon$ , and the other by $g\varepsilon$. The corresponding Hall conductivities, $\sigma(\varepsilon)$ and $\sigma(g\varepsilon)$, transform as $\sigma(g\varepsilon)=g^{-1}\sigma(\varepsilon)$. Since the action of $g$ on $\sigma$ satisfies $g^{-1}\sigma=-\sigma$, it follows that 
\begin{eqnarray}\label{sigma1}
\sigma(g\varepsilon)	=	-\sigma(\varepsilon). 
\end{eqnarray}
Therefore, a sign reversal of the anomalous transport is achieved.

 As an illustration, consider a 2D AM with $\left[C_{2\perp}||C_{4z}\right]$ symmetry, which relates two opposite magnetic atoms, as shown in Fig.~\ref{fig1}(b). This symmetry operation transforms energy eigenvalues according to $\left[C_{2\perp}||C_{4z}\right]E(s,k_{x})=E(-s,k_{y})$, indicating that the bands exhibit the same dispersion along $k_{x}$ and $k_{y}$, but with opposite spin $s$ [see bottom panel of Fig.~\ref{fig1}(b)]. When the Néel vector is perpendicular to the plane and SOC is included, the system will contain the magnetic point-group operation $C_{4z}\mathcal{T}$, which enforces a vanishing anomalous Hall conductivity. 
Furthermore, to obtain an anomalous Hall response, $C_{4}\mathcal{T}$ symmetry relates that magnetic sublattice must be broken, which can be achieved by  introducing a uniaxial strain $\varepsilon_{xx}$  along the $x$ axis. The other equivalent symmetry-broken state is obtained by acting $C_{4z}\mathcal{T}$ on $\varepsilon_{xx}$ , giving $\varepsilon_{yy}= C_{4z}\mathcal{T}\varepsilon_{xx}$. The corresponding band structures reveal that spin-up and spin-down bands are no longer energetically equivalent due to the absence of $\left[C_{2\perp}||C_{4z}\right]$ symmetry [see Fig.~\ref{fig1}(a) and \ref{fig1}(c)]. Crucially, these two symmetry-breaking configurations, $\varepsilon_{xx}$ and $\varepsilon_{yy}$ yield anomalous Hall conductances $\sigma(\varepsilon_{xx})$ and  $\sigma(\varepsilon_{yy})$ with opposite signs. 
This principle is valid for all other operations in Set (\ref{msg3}). As a consequence, for all 2D AMs with an out-of-plane Néel vector described by the spin group symmetry in Set (\ref{ssg11}), the AHE and its sign reversal can also be achieved by breaking the crystal symmetry that connects the two magnetic sublattices. Thus, a scheme for manipulating anomalous transport in 2D AM has been established. 

\begin{figure}[t]
\includegraphics[width=8.4cm]{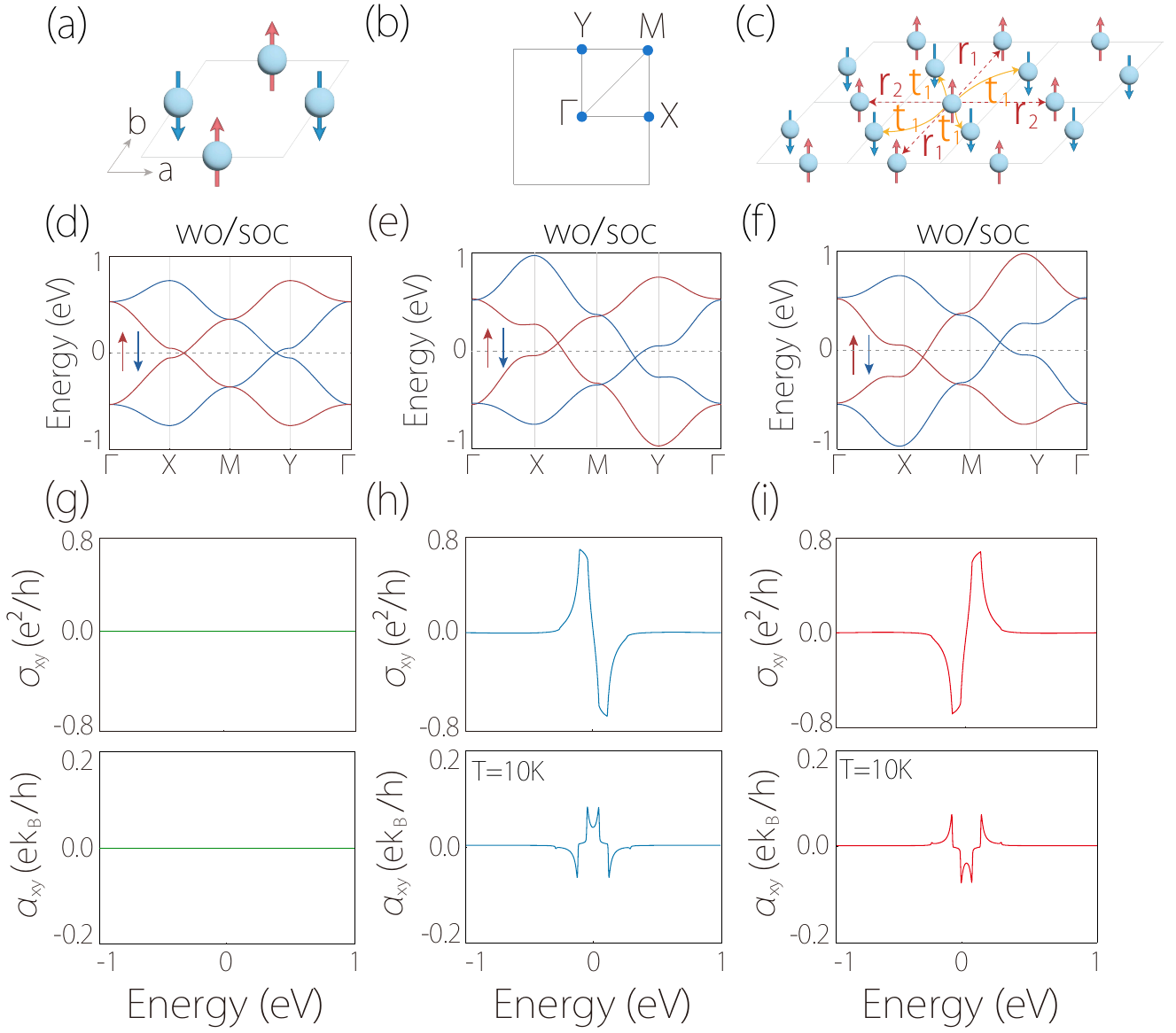}
\caption{(a) The view of the unit cell for the lattice model. (b) Brillouin zone (BZ) of the lattice model. (c) Sketch of the relevant hoppings, with the red arrow and blue arrow representing the two sublattices. (d) Band structure of the model (\ref{tbAM}) along high-symmetry lines without SOC. (e), (f) Energy bands of model (\ref{tbAMbreak1}) and model (\ref{tbAMbreak2}) without SOC. 
(g)-(i) Corresponding energy-dependent anomalous Hall and Nernst conductivities with SOC, calculated from the Hamiltonians $\mathcal{H}(\mathbf{k})$, $\mathcal{H}_{1}(\mathbf{k})$, and $\mathcal{H}_{2}(\mathbf{k})$, respectively.
\label{fig2}}
\end{figure}

We now demonstrate the feasibility of our method by constructing a lattice model based on above symmetry analysis. As an illustrative example, we consider a 2D AM system with $\left[C_{2\perp}||C_{4z}\right]$ symmetry on a square lattice, as shown in Fig.~\ref{fig2}(a). Without loss of generality, a space-group symmetry of No.123 (P4/mmm) is adopted here. Its Brillouin zone (BZ) is shown in Fig.~\ref{fig2}(b). And the unit cell contains two active sites $\left\{ \left(1/2,0,0\right),\left(0,1/2,0\right)\right\} $, which are described by $2f$ Wyckoff position. We set one $\left|s\right\rangle $ orbital at each site. Without altermagnetic ordering, the system present time-reversal symmetry $\mathcal{T}$, and the lattice symmetry can be generated by a four-fold rotation $C_{4z}$, two two-fold rotations: $C_{2z}$, $C_{2x}$, and a spatial inversion symmetry $\mathcal{P}$. Under the basis of $\left|s\right\rangle $ orbital, the matrix representation of these symmetry operators are given by 
\begin{eqnarray}\label{Sym-123}
C_{4z}&=&\sigma_{1},\thinspace\thinspace\thinspace C_{2z}=\sigma_{0},\thinspace\thinspace\thinspace C_{2x}=\sigma_{0}, \nonumber \\
\mathcal{P}	&=&	\sigma_{0},\thinspace\thinspace\thinspace\mathcal{T}=\sigma_{0}\mathcal{K},
\end{eqnarray}
where $\sigma_{i}$ ($i$=1, 2, 3) are Pauli matrices and $\sigma_{0}$ as the $2\times2$ identity matrix. $\mathcal{K}$ is a complex conjugate operator. According to the standard approach \cite{st1, st2}, the symmetry allowed lattice model may be written as \cite{zhang2022magnetictb}.
\begin{eqnarray}\label{tb123}
\mathcal{H}_{0}(\mathbf{k})	& = &	\left[e_{1}+(r_{1}+r_{2})(\mathrm{cos}k_{x}+\mathrm{cos}k_{y})\right]\sigma_{0} \nonumber\\
&  - &  \left[(r_{1}-r_{2})(\mathrm{cos}k_{x}-\mathrm{cos}k_{y})\right]\sigma_{3} \nonumber\\
		&+& 4t_{1}\mathrm{cos}\frac{k_{x}}{2}\mathrm{cos}\frac{k_{y}}{2}\sigma_{1}.
\end{eqnarray}
Here, $e_{1}$ is on-site energy of the model, $t_{1}$ is a nearest-neighbour hopping parameter, and $r_{i}$ are next-nearest-neighbor hopping parameters [see Fig.~\ref{fig2}(c)].

Then, we consider spin degree of freedom and turn
on a AM ordering along $z$-direction [see Fig.~\ref{fig1}(a)]. The AM ordering breaks $\mathcal{T}$, $C_{4z}$ and $C_{2x}$, while holds $C_{2z}$, $\mathcal{P}$ and the combined operation $C_{4z}\mathcal{T}$, $C_{2x}\mathcal{T}$. The AM Hamiltonian up to the leading order may be expressed as
\begin{equation}\label{tbAM}
\mathcal{H}_{0}^{'}(\mathbf{k})	=	s_{0}\mathcal{H}_{0}(\mathbf{k})+J_{z}s_{z}\sigma_{3},
\end{equation}
with $J_{z}$ the strength of the AM potential and $\boldsymbol{s}$ represent the Pauli matrixes acting on spin space. The band structure of this model (\ref{tbAM}) with $e_{1}$=0, $t_{1}$=0.1, $r_{1}$=0.1, $r_{2}$=-0.1, $J_{z}$=0.35 is plotted in Fig.~\ref{fig2}(d) [The units of the hopping
parameters here and below are all eV]. One can find a spin-up Weyl point appearing along the $X$-$M$ path due to the band inversion at the $X$ point. Considering the spatial inversion symmetry, another Weyl point exists on the opposite path, thus forming a pair of Weyl points with the spin-up channel. In addition, opposite spin sublattices in the model (\ref{tbAM}) are related by $C_{4z}\mathcal{T}$ symmetry, thus another pair of Weyl points with the spin-down channel will also appear, as shown in the $Y$-$M$ path of Fig.~\ref{fig2}(d).  The system hosts a pair of spin-up and a pair of spin-down Weyl points lying at the same energy.

To achieve control over anomalous transport properties, the key issue is breaking the symmetry that couples magnetic sublattices, as established by our previous analysis. As illustrated in Fig.~\ref{fig1}, there are two distinct ways for symmetry breaking. Proceeding with this approach, we introduce the first $C_{4z}\mathcal{T}$ symmetry-breaking term, which may be given by

\begin{equation}
H_{break-I}(\mathbf{k})	=	\frac{1}{2}\left(\Delta+2r_{a}\mathrm{cos}k_{x}+2r_{b}\mathrm{cos}k_{y}\right)\left(\Gamma_{0,0}+\Gamma_{0,3}\right)
\end{equation}
with $\Delta$, $r_{a(b)}$ the parameters, and $\Gamma_{i,j}=\sigma_{i}\otimes\sigma_{j}$. The corresponding Hamiltonian is then written as
\begin{equation}\label{tbAMbreak1}
\mathcal{H}_{1}^{'}(\mathbf{k})	=	\mathcal{H}_{0}^{'}(\mathbf{k})+H_{break-I}(\mathbf{k}).
\end{equation}
The band structure of model (\ref{tbAMbreak1}) is presented in Fig.~\ref{fig2}(e) and the parameters of the term $H_{break-I}$ are set as $\Delta$= 0, $r_{a}$=-0.06, $r_{b}$= 0.05. Such a nonzero $H_{break-I}$ breaks the energy equivalence between spin-up and spin-down bands, and spin-up Weyl points locate above the Fermi level, whereas their spin-down counterparts reside below. Additionally, we introduce a second $C_{4z}\mathcal{T}$ symmetry-breaking term that is obtained by a symmetry transformation of the first one, namely
\begin{eqnarray}\label{break2}
H_{break-II}(\mathbf{k})	&=&	AH_{break-I}(-k_{y},k_{x})A^{-1}\nonumber\\
	&=&	\frac{1}{2}\left(\Delta+2r_{b}\mathrm{cos}k_{x}+2r_{a}\mathrm{cos}k_{y}\right) \nonumber\\
		&\times&\left(\Gamma_{0,0}-\Gamma_{0,3}\right)
\end{eqnarray}
with $A=C_{4z}T=\Gamma_{0,1}\mathcal{K}$. Accordingly, the Hamiltonian including the second symmetry-breaking term becomes
\begin{equation}\label{tbAMbreak2}
\mathcal{H}_{2}^{'}(\mathbf{k})	=	\mathcal{H}_{0}^{'}(\mathbf{k})+H_{break-II}(\mathbf{k}).
\end{equation}

The calculated band structure of model (\ref{tbAMbreak2}) is shown in Fig.~\ref{fig2}(f) [using the same parameters as in Fig.~\ref{fig2}(e), which is based on model (\ref{tbAMbreak1})]. It is easy to find that, compared with $H_{break-I}(\mathbf{k})$, $H_{break-II}(\mathbf{k})$ imposes an opposite energy shift on the Weyl points, pushing the spin-up Weyl point below and the spin-down Weyl point above the Fermi level. Beyond the band structure effects, the two symmetry-breaking terms play a more critical role in driving the emergence of anomalous transport and its sign reversal. We then examine the relativistic AHE and ANE with spin-orbit coupling (SOC) included. The symmetry-allowed SOC term is given by
\begin{equation}\label{socterm}
H_{soc}	=	\left(\begin{array}{cccc}
0 & ih_{1} & 0 & 0\\
-ih_{1} & h_{2} & 0 & 0\\
0 & 0 & h_{2}^{'} & -ih_{1}\\
0 & 0 & ih_{1} & 0
\end{array}\right),
\end{equation}
with $h_{1}=4 t_{soc}\mathrm{sin}\frac{k_{x}}{2}\mathrm{sin}\frac{k_{y}}{2}$, $h_{2}=\delta+2r_{c}\mathrm{cos}k_{x}+2r_{d}\mathrm{cos}k_{y}$ and $h_{2}^{'}=\delta+2r_{d}\mathrm{cos}k_{x}+2r_{c}\mathrm{cos}k_{y}$. Here, $t_{soc}$, $\delta$ and $r_{c(d)}$ are real parameters. Finally, the corresponding Hamiltonian with SOC, including two types of $C_{4z}\mathcal{T}$-breaking terms, is given by
\begin{equation}\label{socham12}
\mathcal{H}_{1(2)}(\mathbf{k})	=	\mathcal{H}_{1(2)}^{'}(\mathbf{k})+H_{soc}.
\end{equation}
And the Hamiltonian with SOC in the absence of symmetry-breaking terms reduces to
\begin{equation}\label{socham0}
\mathcal{H}(\mathbf{k})	=	\mathcal{H}_{0}^{'}(\mathbf{k})+H_{soc}.
\end{equation}
As expected, the inclusion of SOC gaps out all these Weyl points. The band structures of the Hamiltonians with SOC, namely $\mathcal{H}(\mathbf{k})$, $\mathcal{H}_{1}(\mathbf{k})$, and $\mathcal{H}_{2}(\mathbf{k})$, are presented in the SM~\cite{sm}, where the SOC parameters are set as $t_{soc}$=0.015,  $r_{c}$= $r_{d}$=0.004 and $\delta$ =0. Accordingly, the anomalous Hall conductivity $\sigma_{xy}$ and the anomalous Nernst conductivity $\alpha_{xy}$, evaluated using these Hamiltonians, are plotted in Figs.~\ref{fig2}(g)-\ref{fig2}(i). One can observe that for pristine AM with an out-of-plane Néel vector, the $\sigma_{xy}$ and $\alpha_{xy}$ are always zero [see Fig.~\ref{fig2}(g)]. This is dictated by the $C_{4}\mathcal{T}$, which forces the Berry curvature contributions from the gapped Weyl points with opposite spins to be equal in magnitude and opposite in sign, resulting in a zero integral over the BZ. For comparison, the corresponding results for $\mathcal{H}_{1}(\mathbf{k})$ and $\mathcal{H}_{2}(\mathbf{k})$ with the same parameters are shown in Figs.~\ref{fig2}(h) and (i), respectively. Clearly, the two symmetry-breaking terms can induce both the AHE and ANE, with opposite signs in the anomalous transport conductivities $\sigma_{xy}$ and $\alpha_{xy}$. This originates from the breaking of the symmetry connecting the two magnetic sublattices, which causes the opposite Berry curvatures in momentum space to be no longer correlated and thus unable to cancel each other, thereby realizing our proposal.

In addition, when symmetry permitting, one can tune the model parameters to leave only single-spin Weyl points near the Fermi level, the quantum anomalous Hall effect (QAHE) and its sign reversal can be realized with SOC included (see SM~\cite{sm}). It should be clarified that realizing the QAHE is highly parameter-dependent, meaning its observation in real materials is not always guaranteed. Conversely, the AHE and its sign reversal proposed above are exceptionally robust, merely requiring the presence of the perturbation term regardless of its magnitude, and thus can be readily realized in real materials.

\section{ Material realizations}
In this section, we demonstrate our idea by first-principles calculations on two concrete materials. We show that applying a finite crystalline symmetry-breaking perturbation to these materials is sufficient to realize the anomalous transport response and its sign reversal. 
\begin{figure}[b]
\includegraphics[width=8.1cm]{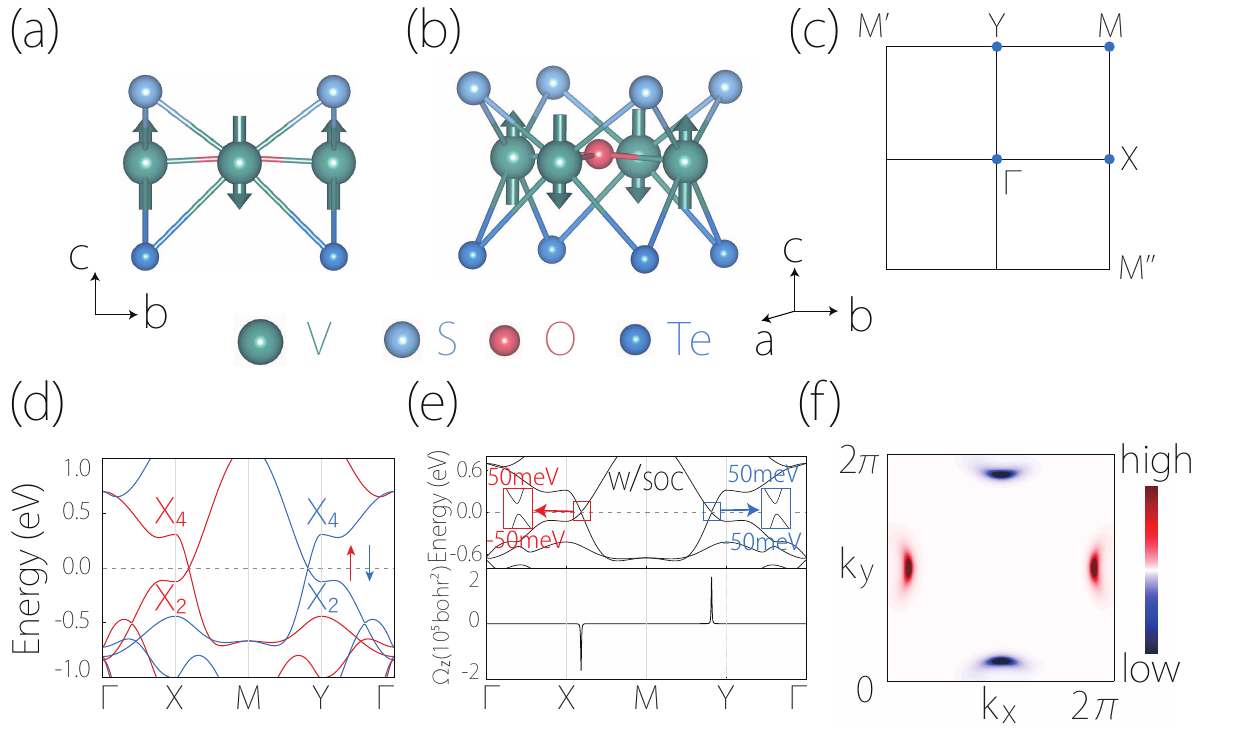}
\caption{(a) Side and (b) front views of the crystal structure of monolayer
V$_{2}$STeO. (c) BZ with high-symmetry points labeled. (d), (e) Electronic band structure of V$_{2}$STeO without (w/o) and with (w) SOC, respectively. In (d), the blue and red lines correspond to the two spin channels, and the irreducible representation at the $X$ point is indicated. (e) Corresponding band structure with SOC, where the bottom panel displays the Berry curvature along the high-symmetry lines. (f) Reciprocal-space distribution of the Berry curvature within the gap.
\label{fig3}}
\end{figure}
\begin{figure*}[t]
\includegraphics[width=16.4cm]{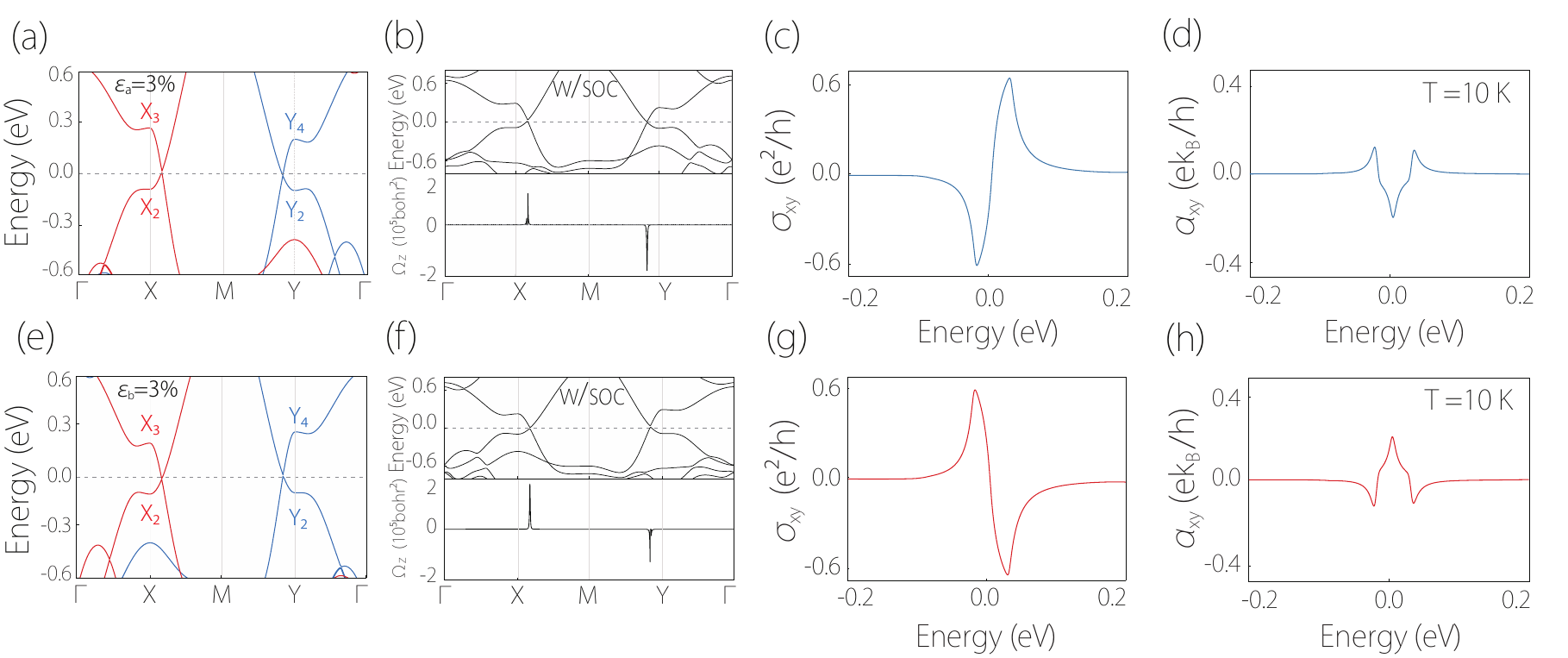}
\caption{(a) Electronic band structure of V$_{2}$STeO without SOC, (b) band structure with SOC and the corresponding Berry curvature, (c) energy-dependent anomalous Hall conductivity, and (d) energy-dependent anomalous Nernst conductivity, all calculated under a 3\% tensile strain along the $a$-axis.  Panels (e)–(h) reproduce the same quantities but under 3\% tensile strain along the $b$-axis.
\label{fig4}}
\end{figure*}

\subsection{Example 1: V$_{2}$STeO}
The first example is monolayer V$_{2}$STeO, which is obtained by substituting Te with S in the top layer of V$_{2}$Te$_{2}$O~\cite{xu2026chemical, li2024strain}.
We note that the parent monolayer V$_{2}$Te$_{2}$O~\cite{PhysRevB.108.024410, PhysRevB.111.184437} also serves as a valid candidate, with its detailed properties presented in the SM~\cite{sm}.
As shown in Figs.~\ref{fig3}(a) and \ref{fig3}(b), the V$_{2}$STeO features a square lattice with the symmorphic space group $P4mm$ (No.99), generated by the symmetry operations $C_{2z}$, $C_{4z}$, and $M_{y}$. The corresponding BZ is shown in Fig.~\ref{fig3}(c). 
Recent investigations identify V$_{2}$Te$_{2}$O as a candidate for altermagnetism, featuring a robust out-of-plane easy axis and verified dynamical and thermodynamical stability \cite{xu2026chemical}. 
Generally, in the absence of SOC, the symmetry of AM material is described by a spin space group, whose operations can generally be written in the form $\left[E||\mathbf{H}\right]+\left[C_{2\perp}||\mathbf{G}-\mathbf{H}\right]$~\cite{PhysRevX.12.031042}. The $\mathbf{G}$  are the crystallographic Laue groups, and $\mathbf{H}$ is a halving subgroup of the $\mathbf{G}$. The group $\left[E||\mathbf{H}\right]$ consists of symmetries that interchange atoms within the same spin sublattice, whereas the coset $\left[C_{2\perp}||\mathbf{G}-\mathbf{H}\right]$ denotes the exchange of atoms between opposite-spin sublattices. For the V$_{2}$STeO, the nonrelativistic symmetry group contains the following symmetry operations

\begin{align}
\left[E||\left\{ E,C_{2z},M_{y},M_{x}\right\} \right]+\left[C_{2\perp}||\left\{ C_{4z}^{+},C_{4z}^{-},M_{xy},M_{x\overline{y}}\right\} \right].
\end{align}
Here, $E$, $C$, and $M$ denote the identity, rotation, and mirror operations, respectively.

The calculated spin-resolved band structure of V$_{2}$STeO without SOC is presented in Fig.~\ref{fig3}(d) (see SM~\cite{sm} for methods). One can observe that V$_{2}$STeO exhibits AM semimetallic behavior, with both spin-up and spin-down channels featuring two Weyl points at the Fermi level. Specifically, in the spin-up channel, the formation of two Weyl points at the high-symmetry line $M$-$X$-$M^{\prime\prime}$ derive from the band inversion of $X$ point with band representations (BRRs) $X_{2}$ and $X_{4}$. These two Weyl points are connected by $\left[E||M_{y}\right]$, and are protected by $\left[E||M_{x}\right]$, as two band forming the Weyl point respectively belong to band representations $\Delta_{1}$ and $\Delta_{2}$, which have opposite eigenvalues of $\left[E||M_{x}\right]$. Owing to the presence of spin group symmetries $\left[C_{2\perp}||C_{4z}\right]$ and $\left[C_{2\perp}||M_{xy/x\overline{y}}\right]$, a similar band inversion occur in $Y$ point, lead to two spin-down Weyl points emerge at $M$-$Y$-$M^{\prime}$ path, are connected to $\left[E||M_{x}\right]$, and are protected by $\left[E||M_{y}\right]$. 

We now turn to focus on relativistic AHE of V$_{2}$STeO. Since the AHE is forbidden in 2D AM system with an out-of-plane Néel vector according to our previous analysis, and this rule also applies to V$_{2}$STeO. This fact can be further verified by magnetic symmetry analysis.
In the presence of SOC, our first-principle calculations show that the Néel vector in V$_{2}$STeO favors an out-of-plane orientation ($\boldsymbol{N}$ $\parallel$ [001])~\cite{sm}, consistent with recent results~\cite{xu2026chemical}. This magnetic configuration is described by the magnetic space group (MSG) $P4'm'm$ (No. 99.165), and the corresponding magnetic point group (MPG) is 
\begin{align}
4'm'm = \{ &E, C_{2z}, M_{xy}, M_{x\overline{y}}, C_{4z}^{+}T, C_{4z}^{-}T,M_{x}T, M_{y}T \}.
\end{align}

Clearly, the operations $C_{4z}T$ and $M_{xy, x\overline{y}}$ enforce the anomalous Hall response of V$_{2}$STeO to vanish, consistent with our prediction. In addition, the electronic band structure of the monolayer V$_{2}$STeO with SOC is shown in top panel of Fig.~\ref{fig3}(e). Notably, the nonrelativistic spin splitting still survives the inclusion of SOC (see the SM~\cite{sm}). Owing to coupling between spin and lattice that breaks the $\{E||M_{x}\}$ and $\{E||M_{y}\}$ symmetry, both spin-up and spin-down Weyl points gap out [see the inset of Fig.~\ref{fig3}(e)] and generate opposite Berry curvatures [see bottom panel of Fig.~\ref{fig3}(e)], whose contributions to the anomalous Hall conductivity cancel each other. The complete distribution of Berry curvature in BZ as shown in Fig.~\ref{fig3}(f). This distribution is related by $C_{4z}\mathcal{T}$ symmetry, and yields a vanishing total flux.

To obtain a finite anomalous Hall response, a perturbation that lowers the symmetry of the system must be introduced. We note that the operations $C_{4z}\mathcal{T}$ and $M_{xy, x\overline{y}}$ suppress the anomalous Hall response while connecting two magnetic sublattices, indicating that breaking the magnetic sublattice symmetry can lead to the emergence of the AHE. We then apply a uniaxial tensile strain of 3\% along the $a$-axis to V$_{2}$STeO. This strain breaks the $C_{4z}$ crystal symmetry, naturally removing the spin group symmetry
$\left[C_{2\perp}||C_{4z}\right]$ and thereby lifting the energetic degeneracy between the spin-up and spin-down bands, as shown in the resulting band structure in Fig.~\ref{fig4}(a). Importantly, as demonstrated in the SM~\cite{sm}, the altermagnetic ground state and out-of-plane easy axis of the system remain robust across the 0\% to 5\% strain range. Moreover, the band inversion at the $X$ and $Y$ points [see BRRs in Fig.~\ref{fig4}(a)], along with the $\left[E||M_{x}\right]$ and $\left[E||M_{y}\right]$ symmetry, is preserved under strain, thereby stabilizing the Weyl points for both spin channels. However, a finite energy splitting arises between the spin-up and spin-down Weyl points [see Fig.~\ref{fig4}(a)]. With the inclusion of SOC, the Weyl points open a tiny gap, yet the system lacks a global band gap and exhibits metallic band characteristics [see top panel of  Fig.~\ref{fig4}(b)]. 
\begin{figure}[b]
\includegraphics[width=8.1cm]{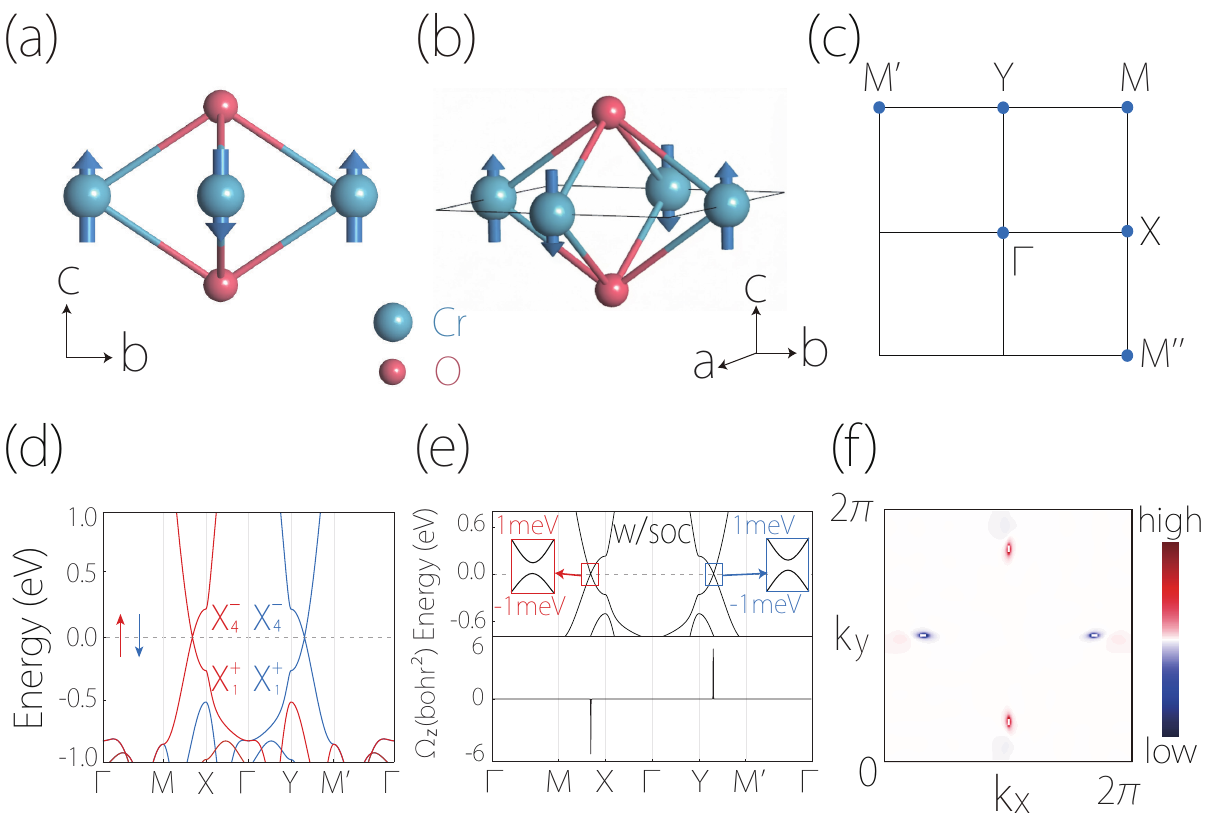}
\caption{(a) Side and (b) front views of the crystal structure of monolayer
Cr$_{2}$O$_{2}$. (c) BZ with high-symmetry points labeled. (d) Spin-resolved band structure without SOC, where blue and red lines denote opposite spin channels, and the irreducible representation at $X$ is marked. (e) Band structure including SOC. The bottom panel plots the Berry curvature along the high-symmetry lines. (f) Reciprocal-space distribution of the Berry curvature within the gap.
\label{fig5}}
\end{figure}
\begin{figure*}[t]
\includegraphics[width=16.8cm]{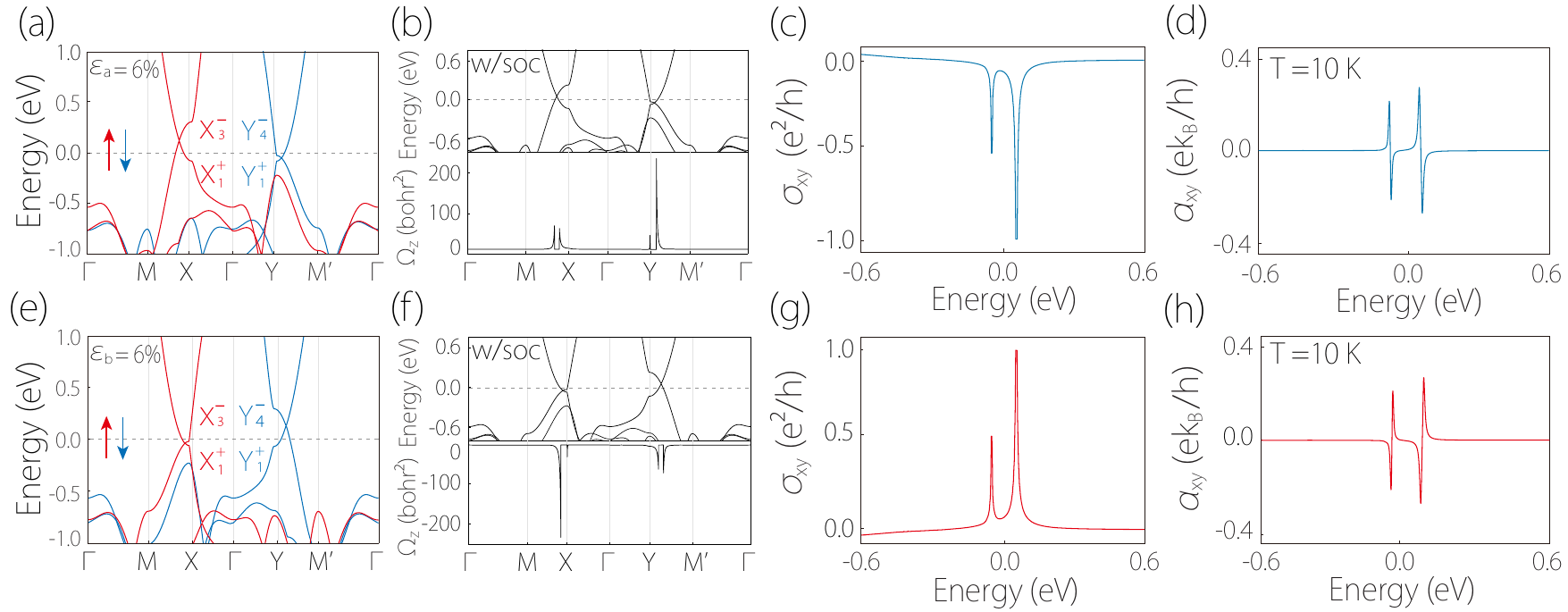}
\caption{Panels (a)–(d) display the band structure of Cr$_{2}$O$_{2}$ without SOC, band structure with SOC and the corresponding Berry curvature, the energy-dependent anomalous Hall conductivity, and the anomalous Nernst conductivity, respectively, under a 6\% uniaxial tensile strain along the $a$-axis. Panels (e)–(h) display the same quantities under a 6\% tensile strain along the $b$-axis.
\label{fig6}}
\end{figure*}

Furthermore, the gapped Weyl points provide the dominant contribution to the Berry curvature of the system, but the contributions from opposite-spin Weyl points can no longer cancel due to the broken symmetry connecting the magnetic sublattices [see bottom panel of Fig.~\ref{fig4}(b)]. As a result, finite anomalous Hall and Nernst responses can emerge, as shown in Figs.~\ref{fig4}(c) and \ref{fig4}(d). On the other hand, the same uniaxial strain can be applied along the $b$-axis to V$_{2}$STeO, with the corresponding band structure shown in Fig.~\ref{fig4}(e). One can observe similar band features with Weyl points  both in opposite spin channels. However, compared to the case of strain along the a-axis, these features exhibit an exchange between the $k_{x}$ and $k_{y}$ directions as well as in the spin degrees of freedom. This characteristic originates from the connection between the two systems, namely V$_{2}$STeO under strain along the $a$-axis and $b$-axis, through the $C_{4z}\mathcal{T}$ symmetry. Moreover, owing to the action of this symmetry, the Berry curvature in the two systems exhibits opposite signs [see bottom panel of Fig.~\ref{fig4}(f)], leading to a sign reversal in both the anomalous Hall conductivity and the anomalous Nernst conductivity [see Figs.~\ref{fig4}(g) and \ref{fig4}(h)]. This enables effective control over the sign of the anomalous Hall transport.

\subsection{Example 2: Cr$_{2}$O$_{2}$}
The second example is monolayer Cr$_{2}$O$_{2}$, which has a square structure with space group $P4/mmm$ (No.123)~\cite{chen2023giant, guo2023quantum}. The symmetry operators of SG 123 are generated by $C_{4z}$, $C_{2y}$ and $P$. The primitive cell contains two Cr and two O atoms [see Figs.~\ref{fig5}(a) and \ref{fig5}(b)], and the corresponding BZ is depicted in Fig.~\ref{fig5}(c). It has been reported that the Cr atoms exhibit antiferromagnetic coupling through O-mediated superexchange~\cite{chen2023giant}. Since the opposite-spin sublattices are connected by rotation symmetry, monolayer Cr$_{2}$O$_{2}$ is classified as an AM. The symmetry operations of the corresponding nontrivial spin group can be expressed as
\begin{align}
&\left[E||\left\{ E,C_{2z},C_{2y},C_{2x},P,M_{z},M_{y},M_{x}\right\} \right] + \notag \\
&\left[C_{2\perp}||\left\{ C_{4z}^{+},C_{4z}^{-},C_{2xy},C_{2x\overline{y}},M_{xy},M_{x\overline{y}},S_{4z}^{+},S_{4z}^{-}\right\} \right].
\end{align}

The calculated spin-resolved band structure for Cr$_{2}$O$_{2}$ is plotted in Fig.~\ref{fig5}(d). One can find that the band of Cr$_{2}$O$_{2}$ is very similar to that of V$_{2}$Te$_{2}$O. Specifically, band inversion at the $X$ ($Y$) forms a pair of spin-up (spin-down) Weyl points along the $M$-$X$-$M^{\prime\prime}$ ($M$-$Y$-$M^{\prime}$) path. Unlike V$_{2}$Te$_{2}$O, the Weyl points in Cr$_{2}$O$_{2}$ feature dual symmetry protection: the Weyl points in the spin-up channel are protected by both $\left[E||C_{2y}\right]$ and $\left[E||M_{x}\right]$ symmetries, whereas the Weyl points in the spin-down channel are protected by $\left[E||C_{2x}\right]$ and $\left[E||M_{y}\right]$. Furthermore, the two magnetic sublattices are connected by $\left[C_{2\perp}||C_{4z}\right]$, $\left[C_{2\perp}||C_{2xy/2x\overline{y}}\right]$, $\left[C_{2\perp}||M_{xy/x\overline{y}}\right]$, and $\left[C_{2\perp}||S_{4z}\right]$, which ensures that the Weyl points in the two spin channels reside at the exact same energy.

Upon including SOC, the spin-lattice coupling pins the Néel vector to a specific direction. Our calculations of the magnetocrystalline anisotropy energy (MAE) indicate an in-plane easy axis for Cr$_{2}$O$_{2}$ (see the SM~\cite{sm} for detailed calculations of the magnetic ground state and MAE), though the energy difference from the out-of-plane configuration is exceptionally small. This result is consistent with previous studies~\cite{chen2023giant}. However, our proposal requires the AM to host an out-of-plane Néel vector to guarantee the emergence of the AHE once the magnetic sublattice symmetry is broken. 
Consequently, the application of an external magnetic field may be tentatively considered as a possible route to reorient the Néel vector of Cr$_{2}$O$_{2}$ out of the plane~\cite{feng2022anomalous}.
The MSG of Cr$_{2}$O$_{2}$ with Néel vector $\boldsymbol{N}$ $\parallel$ [001] belongs to $P4'/mm'm$ (No. 123.342), and corresponding MPG is 
\begin{align}
4'/mm'm = \{ &E, C_{2z}, C_{2xy}, C_{2x\overline{y}}, P, M_{z}, M_{xy}, M_{x\overline{y}}, C_{2x}T,  \notag \\
& C_{2y}T, C_{4z}^{+}T, C_{4z}^{-}T, M_{x}T, M_{y}T, S_{4z}^{+}T, S_{4z}^{-}T \}.
\end{align}
The operations $C_{2xy, 2x\overline{y}}$, $M_{xy, x\overline{y}}$, $C_{4z}T$ and $S_{4z}T$ enforce the anomalous Hall response of Cr$_{2}$O$_{2}$ to vanish while connecting two magnetic sublattices.

Moreover, the band structure of the Cr$_{2}$O$_{2}$ with SOC is shown in top panel of Fig.~\ref{fig5}(e). Owing to the breaking of the $\left[E||C_{2y}\right]$,  $\left[E||M_{x}\right]$, $\left[E||C_{2x}\right]$, and $\left[E||M_{y}\right]$ symmetries under spin-lattice coupling, the Weyl points in both the spin-up and spin-down channels are gapped out, producing opposite Berry curvatures [see bottom panel of Fig.~\ref{fig5}(e)]. Figure~\ref{fig5}(f) shows the complete distribution of the Berry curvature in the BZ. These local contributions are constrained by the symmetries connecting the two magnetic sublattices, leading to an exact cancellation of the Berry curvature over the entire BZ.

Next, to break the symmetry connecting the two magnetic sublattices, we then  apply a uniaxial tensile strain of 6\% along the $a$-axis to Cr$_{2}$O$_{2}$, and the resulting band structure without SOC is shown in Fig.~\ref{fig6}(a). The band inversion at the $X$ and $Y$ points [see BRRs in Fig.~\ref{fig6}(a)], along with the $\left[E||C_{2y}\right]$,  $\left[E||M_{x}\right]$, $\left[E||C_{2x}\right]$, and $\left[E||M_{y}\right]$ symmetry, is preserved under strain, thereby stabilizing the Weyl points for both spin channels. However, a significant energy splitting arises between the spin-up and spin-down Weyl points [see Fig.~\ref{fig6}(a)] owing to the symmetry breaking of $\left[C_{2\perp}||C_{4z}\right]$. With the inclusion of SOC, the Weyl points open a tiny gap, yet the system lacks a global band gap and exhibits metallic band characteristics [see top panel of  Fig.~\ref{fig6}(b)]. Besides, the Berry curvature from opposite-spin Weyl points can no longer cancel due to the broken symmetry connecting the magnetic sublattices [see bottom panel of Fig.~\ref{fig6}(b)]. As a result, finite anomalous Hall and Nernst responses can emerge, as shown in Figs.~\ref{fig6}(c) and \ref{fig6}(d).

Furthermore, applying the same uniaxial strain  along the $b$-axis to Cr$_{2}$O$_{2}$ can also produce similar band characteristics and anomalous transport phenomena. Note that the systems with the same strain applied along the $a$ and $b$ axes are related by $C_{4z}\mathcal{T}$ symmetry. Consequently, the spin and momentum of the energy bands in these two systems are interchanged [see Fig.~\ref{fig6}(a) and \ref{fig6}(e)]. Compared with the $a$-axis case, applying strain along the $b$ axis reverses the sign of the Berry curvature [see the bottom panel of Fig.~\ref{fig6}(f)]. This directly induces a corresponding sign reversal in the anomalous Hall and Nernst conductivities [see Figs.~\ref{fig6}(g) and \ref{fig6}(h)].

\section{CONCLUSION AND DISCUSSION}
In summary, we propose a new strategy to achieve the controllable emergence and sign reversal of the AHE and ANE in 2D AM materials. This strategy only requires the Néel vector of the 2D AM to be out-of-plane, which results in a vanishing anomalous transport signal. Upon applying a perturbation that breaks the crystal symmetry connecting the two magnetic sublattices, the anomalous transport signal can be induced. By constructing a lattice model, we show that the sign of the anomalous transport can be tuned by selecting different perturbation terms, enabling the realization of sign reversal.  
Furthermore, taking V$_{2}$STeO and Cr$_{2}$O$_{2}$ as two material examples, we reveal that anomalous transport signals can be induced by applying a tensile strain along the $a$-axis, which breaks the $C_{4z}\mathcal{T}$ symmetry. Subsequently, when the strain direction is altered to the $b$-axis, the signs of both the anomalous Hall conductance and the anomalous Nernst conductances are reversed. We emphasize that the central contribution of our proposal is the symmetry-controlled tuning of anomalous transport, particularly its sign reversal. By engineering crystal symmetries, we establish a predictive design principle for switching the sign of the anomalous Hall conductivity. Besides, if the system further exhibits multiferroicity~\cite{zhu2025two, camerano2025multiferroic}, such as ferroelasticity~\cite{jql2-b4c4}, the ferroelastic switching corresponds to a lattice rotation that changes the spin-momentum coupling in the altermagnet. Since such ferroelastic switching is highly compatible with the strain-control scheme of the anomalous Hall response, these systems hold promise for achieving the dual-switching of both altermagnetic spin splitting and anomalous transport properties.

\begin{figure}[tb]
\includegraphics[width=8.1cm]{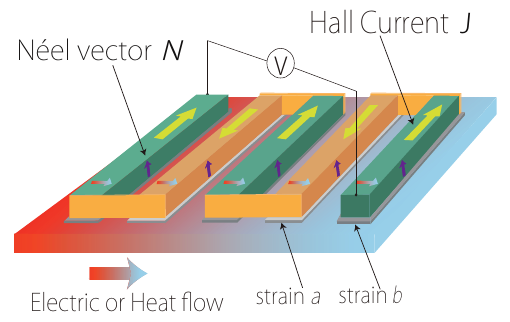}
\caption{Schematic illustration of a cascaded Hall device utilizing the crystal-symmetry-controlled AHE and ANE. The generated voltage polarity can be altered via $a$- or $b$-axis strain, without changing the magnetization of the material.
\label{fig7}}
\end{figure}

Based on the mechanism of crystal-symmetry-controlled anomalous transport, we can offer new design strategies for cascaded Hall devices.
Typically, to enhance the transverse voltage, ferromagnetic or antiferromagnetic materials hosting the ANE (or AHE) are made into a cascaded structure~\cite{Ikhlas2017, guin2019zero}. This allows individual modules with alternating magnetizations to be connected in series laterally. However, a critical issue is precisely achieving the alternating magnetization orientations in adjacent units. For ferromagnets, this requires local writing via external fields, while for antiferromagnets, it remains a major challenge owing to the difficulty in manipulating the Néel vector. Remarkably, the discovery of the crystal-symmetry-controlled AHE and ANE allows us to design a new Hall device without requiring an alternating arrangement of the Néel vector. As shown in Fig.~\ref{fig7}, in contrast to conventional cascaded Hall devices, this structure does not require altering the Néel vector directions (indicated by the red arrows) of the modules. Instead, applying strain to the material (e.g., V$_{2}$STeO) to control the transverse voltage (indicated by the yellow arrows) is sufficient to connect the individual modules in series. Moreover, the strain can be permanently introduced via substrate growth, thereby realizing a highly robust and novel Hall device.

In addition, it is worth noting that our proposed scheme for crystal-symmetry-controlled anomalous transport is universal. In addition to the strain method employed in this work, electric or magnetic fields can also achieve this, as long as these external fields can break the symmetry to connect the magnetic sublattice. Furthermore, the proposed scheme can be generalized to broader classes of 2D antiferromagnets hosting out-of-plane Néel vectors. In these systems, an out-of-plane electric field could serve as an effective means to lift the magnetic sublattice symmetry of antiferromagnets~\cite{PhysRevLett.117.267203}, thus triggering an anomalous Hall response. Our findings open a new avenue for designing next-generation spintronic devices based on anomalous transport in altermagnetic materials.

\begin{acknowledgments}
We thank Jin Cao for helpful discussions. This work is supported by the National Natural Science Foundation of China (Grant No. 12564017, No. 12304086, No. 12304066, Grant No. 12304165), the Basic Research Program of Jiangsu (Grant No. BK20230684).
\end{acknowledgments}

\appendix

\counterwithin{figure}{section}  %
\setcounter{figure}{0}
\renewcommand{\thefigure}{S\arabic{figure}}
%


\bibliography{ref}
\end{document}


\title{Supplementary materials
	 for $``$Manipulating anomalous transport via crystal symmetry in two-dimensional altermagnets$"$}
\author{Dan Li}
\thanks{These authors contributed equally to this work.}
\address{Research Center for Quantum Physics and Technologies, School of Physical Science and Technology, Inner Mongolia University, Hohhot 010021, China}
\address{Key Laboratory of Semiconductor Photovoltaic Technology and Energy Materials at Universities of Inner Mongolia Autonomous Region, Inner Mongolia University, Hohhot 010021, China}

\author{Shuaiyu Wang}
\thanks{These authors contributed equally to this work.}
\address{Research Center for Quantum Physics and Technologies, School of Physical Science and Technology, Inner Mongolia University, Hohhot 010021, China}

\author{Jiabing Chen}
\address{Research Center for Quantum Physics and Technologies, School of Physical Science and Technology, Inner Mongolia University, Hohhot 010021, China}
\address{Key Laboratory of Semiconductor Photovoltaic Technology and Energy Materials at Universities of Inner Mongolia Autonomous Region, Inner Mongolia University, Hohhot 010021, China}

\author{Zeling Li}
\address{Research Center for Quantum Physics and Technologies, School of Physical Science and Technology, Inner Mongolia University, Hohhot 010021, China}
\address{Key Laboratory of Semiconductor Photovoltaic Technology and Energy Materials at Universities of Inner Mongolia Autonomous Region, Inner Mongolia University, Hohhot 010021, China}

\author{Chaoxi Cui}
\email{cuichaoxi@gmail.com}
\affiliation{Key Lab of Advanced Optoelectronic Quantum Architecture and Measurement (MOE), Beijing Key Lab of Nanophotonics \& Ultrafine Optoelectronic Systems, and School of Physics, Beijing Institute of Technology, Beijing 100081, China}

\author{Lei Li}
\address{Research Center for Quantum Physics and Technologies, School of Physical Science and Technology, Inner Mongolia University, Hohhot 010021, China}

\author{Lei Wang}
\address{Research Center for Quantum Physics and Technologies, School of Physical Science and Technology, Inner Mongolia University, Hohhot 010021, China}
\address{Inner Mongolia Key Laboratory of Microscale Physics and Atom Innovation, Inner Mongolia University, Hohhot 010021, China}

\author{Zhi-Ming Yu}
\affiliation{Key Lab of Advanced Optoelectronic Quantum Architecture and Measurement (MOE), Beijing Key Lab of Nanophotonics \& Ultrafine Optoelectronic Systems, and School of Physics, Beijing Institute of Technology, Beijing 100081, China}

\author{Xiaodong Zhou}
\email{zhouxiaodong@tiangong.edu.cn}
\address{School of Physical Science and Technology, Tiangong University, Tianjin 300387, China}

\author{Xiao-Ping Li}
\email{xpli@imu.edu.cn}
\address{Research Center for Quantum Physics and Technologies, School of Physical Science and Technology, Inner Mongolia University, Hohhot 010021, China}
\address{Key Laboratory of Semiconductor Photovoltaic Technology and Energy Materials at Universities of Inner Mongolia Autonomous Region, Inner Mongolia University, Hohhot 010021, China}

\maketitle

\section{symmetry analysis and lattice model calculations}
\subsection{Reduction of spin space group}
In this section, we provide a detailed analysis of the reduction from the spin group to the magnetic group. Furthermore, we present the band structure of the lattice model with a symmetry-breaking term in the presence of spin-orbit coupling (SOC), and demonstrate the sign reversal of the quantum anomalous Hall effect (QAHE).

For a two-dimensional altermagnet, the general symmetry operations that connect the two spin sublattices in the spin space group can be written as

\begin{equation}\label{ssg1}
\mathbb{G}	\otimes	\mathbb{S},
\end{equation}
with 
\begin{eqnarray}\label{ssg11}
\mathbb{G}	&=&	\bigl\{\left[C_{2\perp(s)}||C_{2\parallel(l)}\right],\left[C_{2\perp(s)}||C_{nz}\right],\left[C_{2\perp(s)}||M_{\parallel(l)}\right], \left[C_{2\perp(s)}||S_{nz}\right]\bigl\}
\end{eqnarray}
and
\begin{eqnarray}\label{ssg12}
\mathbb{S} & = & \left\{ \left[E||E\right],\left[C_{2\perp(s)}||E\right]\mathcal{T},\left[C_{\infty\parallel(s)}||E\right]\right\},
\end{eqnarray}
where the terms in the set $\mathbb{G}$ represent the non-trivial spin-group symmetry depending on magnetic structures and crystal, while the $\mathbb{S}$ represent the spin-only group common to collinear magnetic structures. Specifically, the spin group symmetries take the form of $\left[O_{1}||O_{2}\right]$, where $O_{1}$ and $O_{2}$ symmetry operations act on the decoupled spin and lattice space. The subscripts $\perp(s)$ [$\parallel(s)$] and $\parallel(l)$ [$\perp(l)$] denote directions perpendicular [parallel] to the spin axis in spin space and parallel [perpendicular] to the two-dimensional plane in lattice space, respectively. For instance, $C_{2\perp(s)}$ denotes a two-fold rotation around the axis perpendicular to the collinear spins, $C_{\infty\parallel(s)}$ represents an arbitrary rotation around the collinear spin axis, and $C_{2\parallel(l)}$ signifies a two-fold in-plane rotation on lattice plane.
Furthermore, the symmetry operations in the set $\mathbb{G}$ can be combined with elements of the spin only group $\mathbb{S}$ to produce new forms without changing the physics. For instance, $\left[C_{2\perp(s)}||C_{nz}\right]$ in $\mathbb{G}$ can be transformed into $\left[C_{n\parallel(s)}||C_{nz}\right]\mathcal{T}$ by the following combination:
\begin{eqnarray}\label{ssg21}
\left[C_{n\parallel(s)}||C_{nz}\right]\mathcal{T}	&=&	\left[C_{2\perp(s)}||C_{nz}\right]\ast\left[C_{2\perp(s)}||E\right]\mathcal{T} *\left[C_{\infty\parallel(s)}||E\right]. 
\end{eqnarray}
Here, $\mathcal{T}$ is the time-reversal symmetry, and $C_{\infty\parallel(s)}$ is taken to $C_{n\parallel(s)}$. Applying the same procedure to $\left[C_{2\perp(s)}||S_{nz}\right]$ yields $\left[C_{n\parallel(s)}||S_{nz}\right]\mathcal{T}$. As a result, we obtain a new reorganized set $\mathbb{G}^{'}$, in which
\begin{eqnarray}\label{ssg2}
\mathbb{G}^{'}	&=&	\bigl\{\left[C_{2\perp(s)}||C_{2\parallel(l)}\right],\left[C_{n\parallel(s)}||C_{nz}\right]\mathcal{T},\left[C_{2\perp(s)}||M_{\parallel(l)}\right], \left[C_{n\parallel(s)}||S_{nz}\right]\mathcal{T}\bigl\}.
\end{eqnarray}
Moreover, we consider the case where the Néel vector of the 2D AM is oriented out-of-plane, specifically along the $z$-axis (denoting a coupling between spin and lattice degrees of freedom). Notably, the directions indicated  by $\perp(s)$ in spin space and $\parallel(l)$ in lattice space are the same, which we denote uniformly as $\parallel$, while the $\parallel(s)$ represents the out-of-plane ($z$) direction. Accordingly, with consideration of the out-of-plane Néel vector, the set of non-trivial spin group operations reduces to 
\begin{eqnarray}\label{ssg3}
\mathbb{G}^{''}	&=&	\bigl\{\left[C_{2\parallel}||C_{2\parallel}\right],\left[C_{nz}||C_{nz}\right]\mathcal{T},\left[C_{2\parallel}||M_{\parallel}\right], \left[C_{nz}||S_{nz}\right]\mathcal{T}\bigl\}.
\end{eqnarray}
Owing to the coupling between spin and lattice, the symmetry with considering spin-orbit coupling (SOC) is described by magnetic space groups, which are subgroups of space groups. The elements of magnetic groups are composed of those elements in spin groups for which the operations in spin space and lattice space are identical. Therefore, upon including SOC, the set of spin group operations $\mathbb{G}^{''}\otimes\mathbb{S}$ reduces to the following magnetic group operations:
\begin{eqnarray}\label{msg3}
\left\{ C_{2\parallel},C_{nz}\mathcal{T},M_{\parallel},S_{nz}\mathcal{T}\right\}. 
\end{eqnarray}
%

\begin{figure}[t]
\centering
\includegraphics[width=0.85\columnwidth]{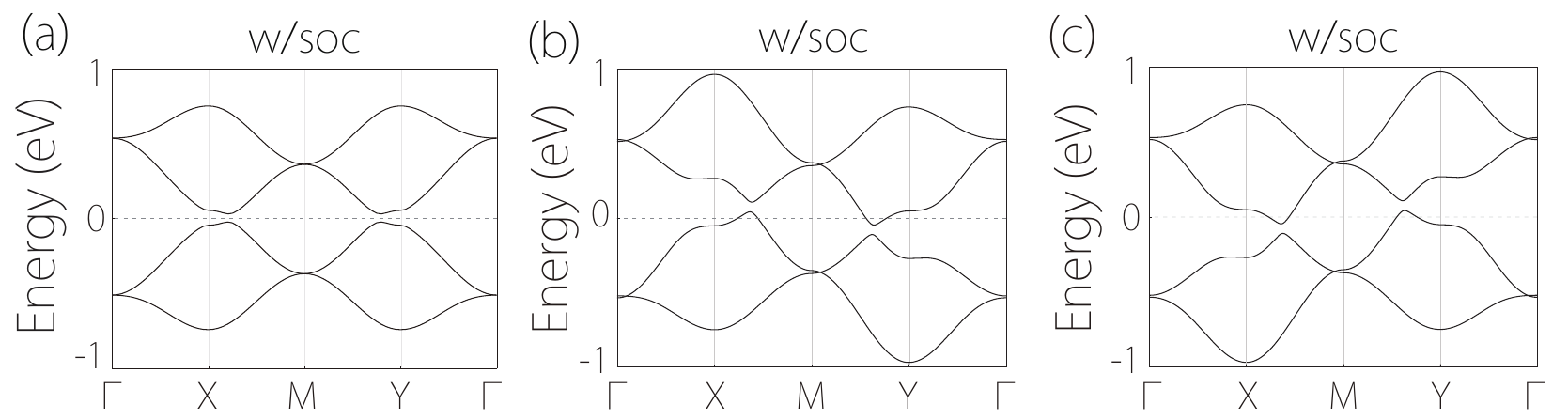} \caption{(a) Band structure of the lattice model with SOC, based on the Hamiltonian $\mathcal{H}(\mathbf{k})$ defined in the main text. (b), (c) Band structures with SOC under different symmetry-breaking terms, derived from $\mathcal{H}_{1}(\mathbf{k})$ and $\mathcal{H}_{2}(\mathbf{k})$, respectively. 
~\label{figs1}}
\end{figure}

In the main text, the band structures without SOC are shown in Figs. 2(d)-2(f) for the lattice model. Here we present the corresponding band structures with SOC. Fig.~\ref{figs1}(a)-(c) show the resulting bands based on the Hamiltonians $\mathcal{H}(\mathbf{k})$, $\mathcal{H}_{1}(\mathbf{k})$, and $\mathcal{H}_{2}(\mathbf{k})$  in the main text, respectively. The SOC parameters are set to $t_{soc}=0.015$, $r_{c}=r_{d}=0.004$, and $\delta=0$. 
\begin{figure}[h]
\centering
\includegraphics[width=0.7\columnwidth]{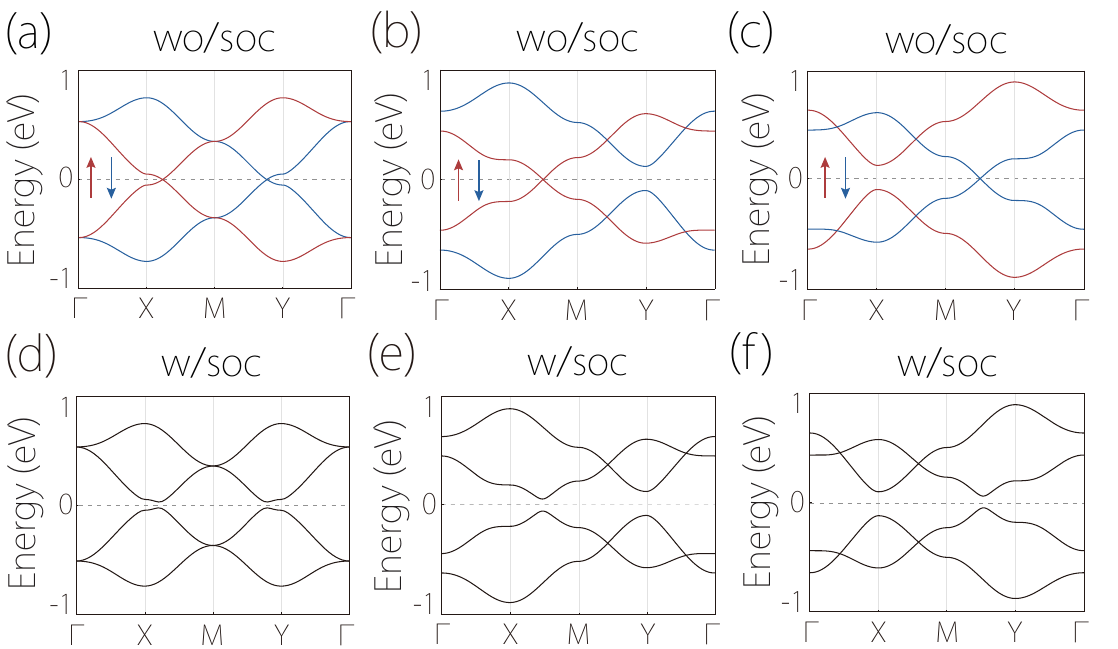} \caption{(a) Band structure without SOC along high-symmetry lines, based on the Hamiltonian $\mathcal{H}_{0}^{'}(\mathbf{k})$. (b), (c) Energy bands of the Hamiltonians $\mathcal{H}_{1}^{'}(\mathbf{k})$ and $\mathcal{H}_{2}^{'}(\mathbf{k})$ without SOC, respectively. (d)-(f) Band structures with SOC corresponding to panels (a)-(c), calculated using $\mathcal{H}(\mathbf{k})$, $\mathcal{H}_{1}(\mathbf{k})$, and $\mathcal{H}_{2}(\mathbf{k})$, respectively. (All Hamiltonians are defined in the main text.)
~\label{figs2}}
\end{figure}

\subsection{Quantum anomalous Hall effect}
Moreover, when symmetry permits, we can further achieve the quantum anomalous Hall effect (QAHE) with its sign reversal. The parameters of the Hamiltonian $\mathcal{H}_{1}^{'}(\mathbf{k})$ are set as $e_{1}$=-0.15, $t_{1}$=0.1, $r_{1}$=0.1, $r_{2}$=-0.1, $J_{z}$=0.35, $\Delta$= 0.3, $r_{a}$=0, and $r_{b}$= -0.01 (Unless stated otherwise, all Hamiltonian expressions hereafter are those given in the main text), and the energy band is plotted in Fig.~\ref{figs2}(b). One can find that there is only a pair of Weyl points in the spin-up channel, which provides a topologically non-trivial band structure for the realization of the QAHE. Furthermore, we can obtain another pair of Weyl points in the spin-down channel using Hamiltonian $\mathcal{H}_{2}^{'}(\mathbf{k})$ with same parameters, as shown in Fig.~\ref{figs2}(c). As a comparison, the bands of the pristine altermagnet, obtained using the Hamiltonian $\mathcal{H}_{0}^{'}(\mathbf{k})$, are shown in Fig.~\ref{figs2}(a). The model parameters are set to $e_{1}$=0, $t_{1}$=0.1, $r_{1}$=0.1, $r_{2}$=-0.1, $J_{z}$=0.35. 

We then consider the SOC effect to open a gap at the Weyl points and achieve quantized anomalous transport properties. 
Further, we plot the band structure with SOC based on Hamiltonian $\mathcal{H}(\mathbf{k})$, $\mathcal{H}_{1}(\mathbf{k})$, and $\mathcal{H}_{2}(\mathbf{k})$, and obtained Figs.~\ref{figs2}(d),~\ref{figs2}(e), and~\ref{figs2}(f), respectively. Here, the SOC-related parameters are $t_{soc}$=0.02,  $r_{c}$= $r_{d}$=0.004 and $\delta$ =0. 
As an result, one can observe that all band crossing points have opened a gap [see Fig.~\ref{figs2}(d)-(f)]. Furthermore, the calculated the Chern number of Fig.~\ref{figs2}(d)-(f), with the results being $0$, $-1$ and $+1$, respectively. Consequently, the distinct topological numbers of $-1$ and $+1$ will naturally lead to anomalous Hall conductivity with opposite signs, thereby enabling control of the quantized anomalous transport via crystal symmetry. 
\begin{figure}[h]
\centering
\includegraphics[width=0.8\columnwidth]{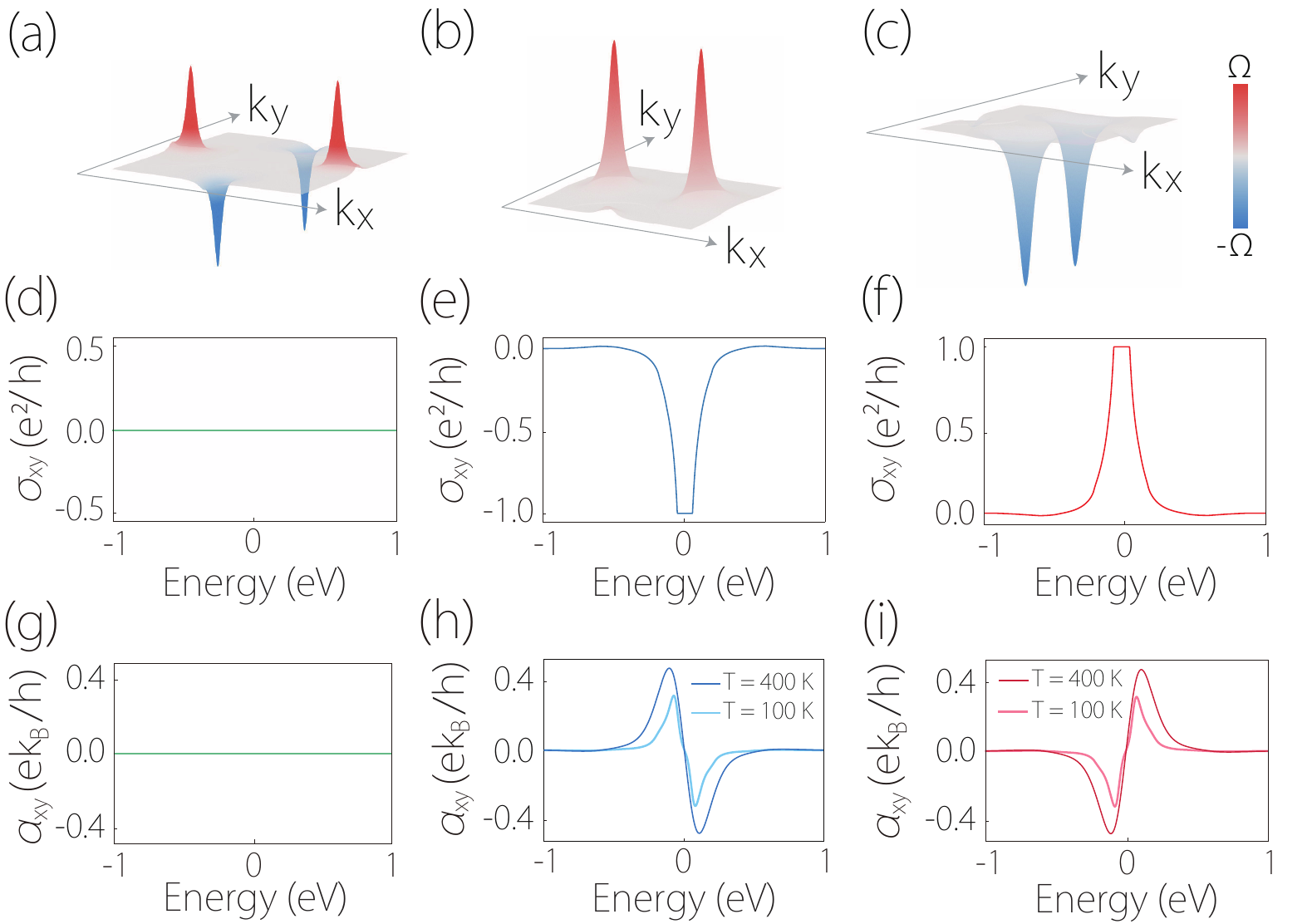} \caption{(a)-(c) The distribution of Berry curvature of the gapped bands for  $\mathcal{H}(\mathbf{k})$, $\mathcal{H}_{1}(\mathbf{k})$, and $\mathcal{H}_{2}(\mathbf{k})$, respectively. (d)-(f) illustrate the energy-dependent anomalous Hall conductivity, while (g)-(i) present the corresponding anomalous Nernst conductivity. 
~\label{figs3}}
\end{figure}
To verify this, we plot the Berry curvature distributions in the Brillouin zone (BZ) for the three phases with Chern numbers $C$= $0$, $-1$, and $+1$, as shown in Fig.~\ref{figs3}(a)-(c). The distribution shows that the main contribution of Berry curvature comes from the gapped Weyl points. In the trivial phase with $C=0$, four pronounced peaks emerge, which can be categorized into two sets with opposite values that satisfy $C_{4z}\mathcal{T}$ symmetry [see Fig.~\ref{figs3}(a)], leading to a vanishing Chern number. For the phase with $C=-1$ ($C=+1$), the Berry curvature exhibits two main positive (negative) peaks of equal magnitude [see Fig.~\ref{figs3}(b) and (c)], corresponding to a pair of massive Weyl points with same spin. Furthermore, by integrating the Berry curvature, we evaluated the intrinsic anomalous Hall conductivity $\sigma_{xy}$ and anomalous Nernst conductivity $\alpha_{xy}$. Fig.~\ref{figs3}(d)-(f) displays the $\sigma_{xy}$ as a function of the energy. The results show that for pristine AM with an out-of-plane Néel vector, the anomalous Hall conductivity is always zero [see Fig.~\ref{figs3}(d)]. Applying two symmetry-breaking perturbations can induce two opposite anomalous Hall conductivities [see Fig.~\ref{figs3}(e) and (f)], and the quantized plateau within the bandgap indicates the characteristic of the QAHE. Moreover, the anomalous Nernst conductivity also exhibits similar characteristics, and the breaking of crystal symmetry can similarly realize the sign reversal of the ANE in our model [see Fig.~\ref{figs3}(g)-(i)].

\section{Details of first-principles calculations}
\subsection{Computational methods}

\begin{figure}[b]
\centering
\includegraphics[width=0.8\columnwidth]{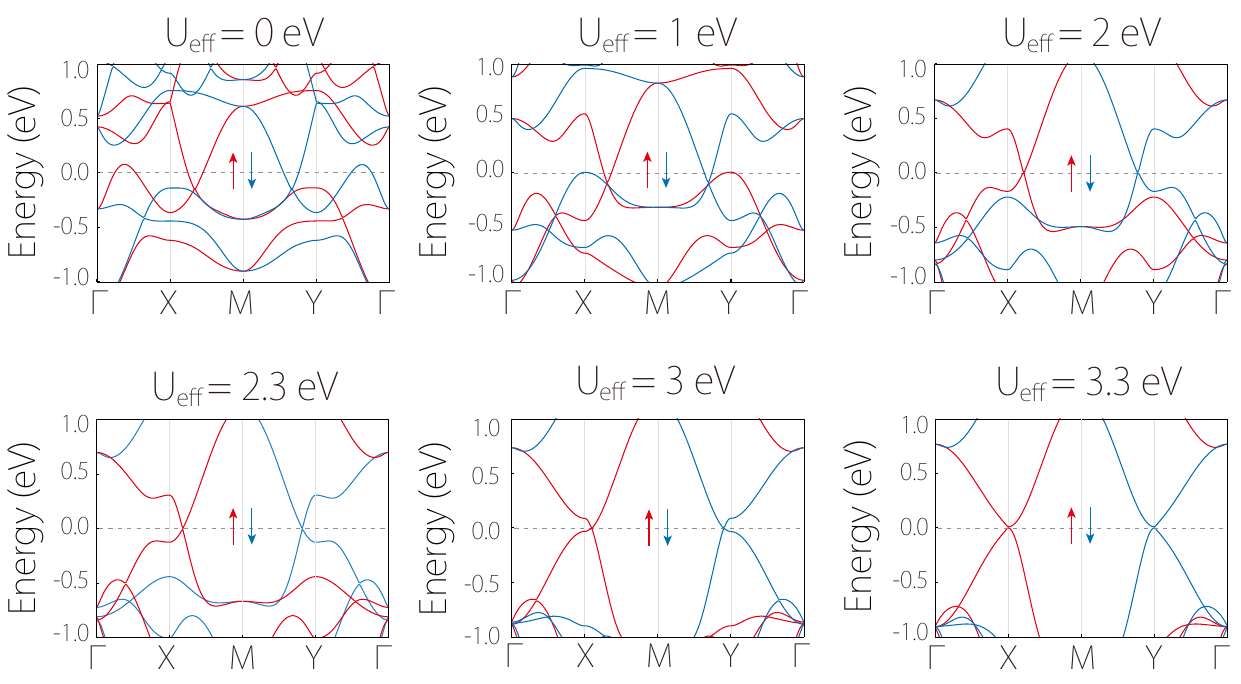} \caption{Spin-resolved band structures of monolayer V$_{2}$STeO with different U values. 
~\label{figs4}}
\end{figure}

The first-principles calculations were performed using the Vienna
\textit{ab initio} Simulation Package (VASP)~\cite{PhysRevB.49.14251, PhysRevB.54.11169} with the projector-augmented wave (PAW) method~\cite{PhysRevB.50.17953}. The exchange-correlation functional was treated within the local spin density approximation (LSDA)~\cite{PhysRevB.23.5048}, with a Hubbard U correction~\cite{PhysRevB.44.943} to account for correlation effects. The Hubbard U values were chosen as 2.3~eV for the V-$d$ orbitals of V$_{2}$STeO, 3.0~eV for the V-$d$ orbitals of V$_{2}$Te$_{2}$O, and 3.55~eV for the Cr-$d$ orbitals of Cr$_{2}$O$_{2}$. The calculated band structures for V$_{2}$STeO with other choices of the Hubbard U are presented in Fig.~\ref{figs4}. The cutoff energies were set to 500 eV for V$_{2}$STeO and V$_{2}$Te$_{2}$O, and 520 eV for Cr$_{2}$O$_{2}$. A $\Gamma$-centered \emph{k}-mesh of $10\times10\times1$, $16\times16\times1$, and $10\times10\times1$ was used for V$_{2}$STeO, Cr$_{2}$O$_{2}$, and V$_{2}$Te$_{2}$O, respectively, for Brillouin zone sampling. The energy and force convergence criteria were set to 10$^{-6}$ eV and 10$^{-2}$ eV/$\mathring{\mathrm{A}}$ for V$_{2}$STeO and V$_{2}$Te$_{2}$O, and 10$^{-7}$ eV and 10$^{-3}$ eV/$\mathring{\mathrm{A}}$ for Cr$_{2}$O$_{2}$. A vacuum layer with a thickness of 20 $\mathrm{\AA}$ was applied to avoid artificial interactions between periodic images. The Berry curvature and intrinsic anomalous Hall conductivity were evaluated using the WANNIER90 package~\cite{mostofi2014updated, PhysRevB.74.195118}.

The intrinsic anomalous Hall conductivity were calculated via the Kubo formula~\cite{PhysRevLett.92.037204, PhysRevLett.97.026603},
\begin{equation}\label{hall}
\sigma_{xy}	=	-\frac{e^{2}}{\hbar}\sum_{n}\int\frac{d^{2}k}{\left(2\pi\right)^{2}}\Omega_{xy}^{n}\left(\mathbf{k}\right)f_{n}\left(\mathbf{k}\right),
\end{equation}
where $\Omega_{xy}^{n}$ is the band- and momentum-resolved Berry curvature
\begin{equation}\label{omega}
\Omega_{xy}^{n}	\left(\mathbf{k}\right)=	-\sum_{n'\neq n}\frac{2\mathrm{Im}\left[\left\langle \psi_{n\mathbf{k}}|\hat{v}_{x}|\psi_{n'\mathbf{k}}\right\rangle \left\langle \psi_{n'\mathbf{k}}|\hat{v}_{y}|\psi_{n\mathbf{k}}\right\rangle \right]}{\left(\omega_{n'\mathbf{k}}-\omega_{n\mathbf{k}}\right)^{2}},
\end{equation}
and $f_{n}\left(\mathbf{k}\right)=1/\left[\mathrm{exp}[(E_{n\mathbf{k}}-\mu)/k_{B}T]+1\right]$ is the Fermi-Dirac distribution function. \(\hat{v}_{x,y}\) represents the velocity operators and \(\psi_{n\mathbf{k}}\) (\(E_{n\mathbf{k}} = \hbar\omega_{n\mathbf{k}}\)) the eigenvector (eigenvalue) at band \(n\) and momentum \(\mathbf{k}\). Besides, the notation $\{x,y\}$, $T$, $\mu$, and $k_{B}$ denote the Cartesian coordinates, temperature, chemical potential, and Boltzmann constant, respectively. As the thermoelectric counterpart of the AHE, the anomalous Nernst conductivity is given by
\begin{equation}\label{ane}
\alpha_{xy}	=	\frac{ek_{B}}{\hbar}\sum_{n}\int\frac{d^{2}k}{\left(2\pi\right)^{2}}\Omega_{xy}^{n}\left(\mathbf{k}\right)S^{n}\left(\mathbf{k}\right)
\end{equation}
with
\begin{eqnarray}\label{s}
S^{n}(\mathbf{k})
&=& \bigl[k_{B}T\ln\bigl(1+\mathrm{e}^{-(E_{n\mathbf{k}}-\mu)/k_{B}T}\bigr) \nonumber\\
&& {}+ (E_{n\mathbf{k}}-\mu)f_{n}(\mathbf{k})\bigr]\big/k_{B}T.
\end{eqnarray}
Here, $S^{n}(\mathbf{k})$ stands for the entropy density.

\begin{figure}[t]
\centering
\includegraphics[width=1.0\columnwidth]{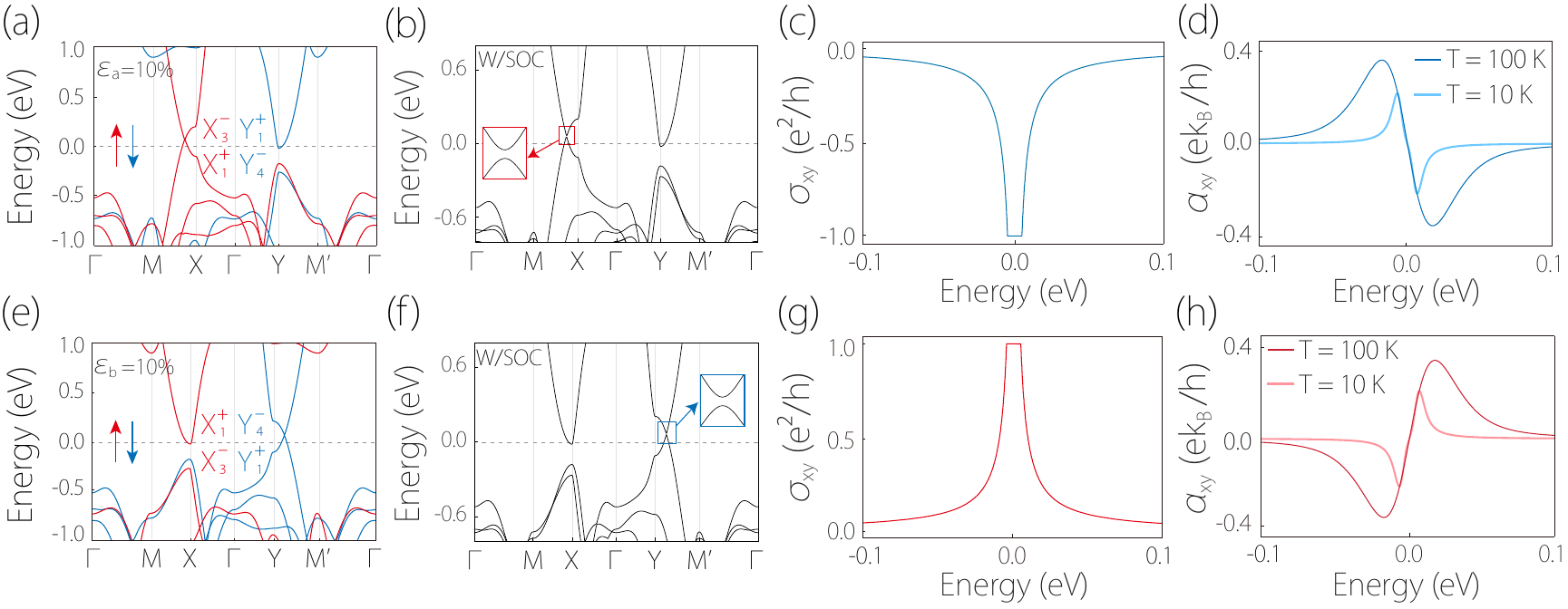} \caption{Electronic band structures of Cr$_{2}$O$_{2}$ (a) without SOC , (b) with SOC, (c) energy-resolved anomalous Hall conductivity, and (d) energy-resolved anomalous Nernst conductivity, all under 10\% tensile strain along the $a$-axis. (e)–(h) reproduce the same quantities but under 10\% tensile strain along the $b$-axis. In (c) and (g), a quantized plateau of the anomalous Hall conductivity emerges within the insulating region.  
~\label{figs5}}
\end{figure}

In addition, we briefly show the emergence of the QAHE and the possibility of its sign reversal in Cr$_{2}$O$_{2}$. Upon further increasing the strain to 10\%, the band structure of Cr$_{2}$O$_{2}$ is shown in Fig.~\ref{figs5}(a). Notably, we observe the band inversion at the $X$ point persists, while that at the $Y$ point vanishes. This leads to the annihilation of a pair of spin-down Weyl points along the $M$-$Y$-$M^{\prime}$ path, leaving only a pair of spin-up Weyl points along the $M$-$X$-$M^{\prime\prime}$ path. With the inclusion of SOC, this pair of Weyl points opens a non-trivial band gap, transforming the system into a Chern insulator with Chern number $C = -1$. Moreover, each gapped Weyl point contributes the Hall conductance with $-0.5$ $e^{2}/h$, resulting in a total Hall conductance as $-1$ $e^{2}/h$, and quantized Hall plateau as shown in Fig.~\ref{figs5}(c). Furthermore, we plot the energy dependence of the anomalous Nernst conductivity, as shown in Fig.~\ref{figs5}(d). Notably, anomalous Nernst conductivity $\alpha_{xy}$ and anomalous Hall conductivity $\sigma{}_{xy}$ are intimately related to each other through the generalized Mott formula \cite{PhysRevLett.97.026603}. In the low-temperature limit, $\alpha_{xy}$ is proportional to the negative energy derivative of $\sigma{}_{xy}$. As a result, we observe that when the energy changes, the slope of $\sigma{}_{xy}$ starts out negative, corresponding to a positive value of‌ $\alpha_{xy}$. After passing through the Hall plateau, the slope of‌ $\sigma{}_{xy}$ ‌becomes positive, while $\alpha_{xy}$ ‌turns negative. The realization of sign reversal in the QAHE follows a scheme analogous to the sign flip of the AHE discussed in the main text. As a contrast, by applying 10\% biaxial strain along the $b$-axis in Cr$_{2}$O$_{2}$, we observe that only the band inversion at $Y$-point remains in the band, resulting in a pair of spin-down Weyl points. With SOC, the system evolves into a Chern insulator with a Chern number of $C = +1$. This leads to a sign reversal in the anomalous Hall conductivity plateau, accompanied by a corresponding reversal in the anomalous Nernst conductivity. 

%

\newpage
\subsection{Spin-projected band structure}
\begin{figure}[h]
\centering
\includegraphics[width=0.8\columnwidth]{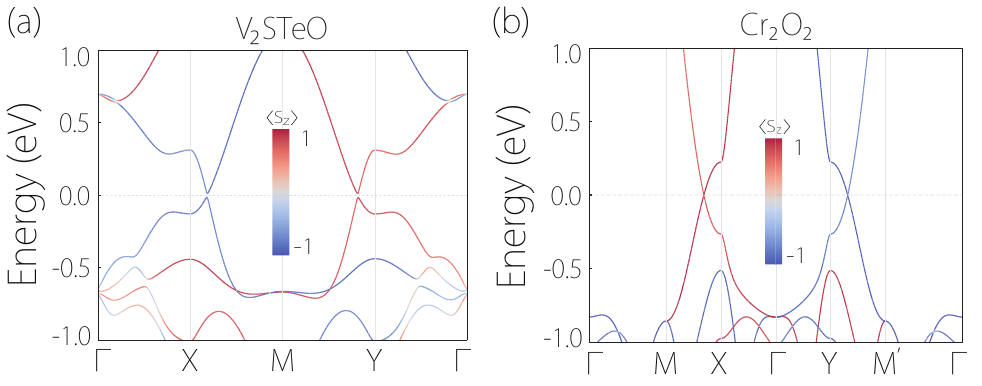} \caption{Spin-projected band structures of (a) V$_{2}$STeO and (b) Cr$_{2}$O$_{2}$ with SOC. The color bar represents the expectation value of $s_{z}$. 
 ~\label{figs6}}
\end{figure}

To explicitly examine the persistence of the nonrelativistic spin splitting in the presence of SOC, we calculate the spin-projected band structures for both candidate materials. Figure~\ref{figs6} shows the resulting band structures of V$_{2}$STeO and Cr$_{2}$O$_{2}$ with SOC taken into account. The color scale denotes the expectation value of the spin operator $s_{z}$, which exhibits a characteristic momentum dependence. Notably, the nonrelativistic spin splitting is preserved upon the inclusion of SOC, which maintains a prominent spin-resolved feature.

%
\subsection{Other candidate: V$_{2}$Te$_{2}$O}
\begin{figure}[t]
\includegraphics[width=0.6\columnwidth]{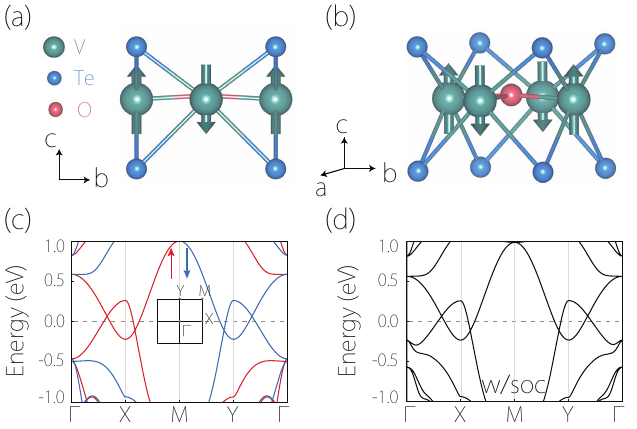} \caption{(a) Side and (b) perspective views of the crystal structure of monolayer V$_2$Te$_2$O. (c) Spin-resolved electronic band structure calculated without SOC. The inset shows the two-dimensional Brillouin zone. (d) Relativistic band structure calculated with SOC.
~\label{figs7}}
\end{figure}

As another material realization, monolayer V$_{2}$Te$_{2}$O crystallizes in a square lattice and hosts a collinear altermagnetic configuration, as shown in Figs.~\ref{figs7}(a) and \ref{figs7}(b). In the nonrelativistic limit, its spin symmetry contains $\left[C_{2\perp}||C_{4z}\right]$, which connects the two oppositely polarized V sublattices and imposes $E(s,k_{x})=E(-s,k_{y})$. Consequently, the spin-up and spin-down bands interchange their dispersions between the $\Gamma$-$X$ and $\Gamma$-$Y$ directions, giving rise to the characteristic $d$-wave spin splitting displayed in Fig.~\ref{figs7}(c). Meanwhile, $[C_{2\perp}\Vert C_{2[110]}]$ and $[C_{2\perp}\Vert C_{2[1\bar{1}0]}]$ enforce spin degeneracy along the two diagonal directions. The band crossings near X and Y originate from two oppositely spin-polarized nodal loops protected by the horizontal mirror symmetry $M_z$. When SOC is included [Fig.~\ref{figs7}(d)], the out-of-plane magnetic configuration belongs to the magnetic layer group $p4^\prime/mm^\prime m$. Although the altermagnetic spin splitting remains visible and the $M_z$-protected crossings survive, the $C_{4z}\mathcal{T}$ operation and the diagonal twofold rotations reverse $\sigma_{xy}$, thereby enforcing a vanishing anomalous Hall conductivity in pristine V$_2$Te$_2$O.

\begin{figure}[h]
\centering
\includegraphics[width=1.0\columnwidth]{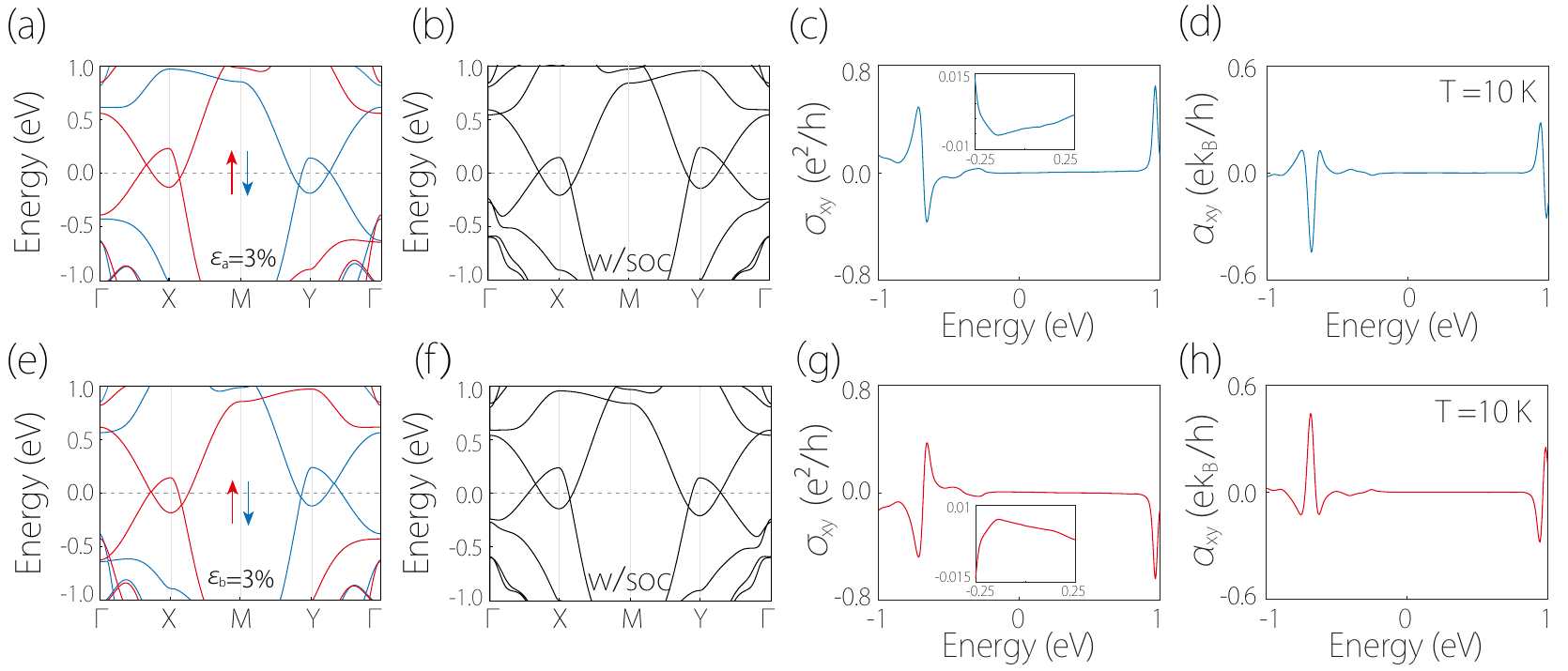} \caption{Electronic structures and anomalous transport properties of monolayer V$_2$Te$_2$O under uniaxial tensile strain. (a) Spin-resolved band structure without SOC and (b) relativistic band structure with SOC under 3\% strain along the $a$ axis. (c) Energy-dependent anomalous Hall conductivity $\sigma_{xy}$ and (d) anomalous Nernst conductivity $\alpha_{xy}$ at $T=10$~K for the $a$-strained configuration. Panels (e)-(h) show the corresponding results under 3\% strain along the $b$ axis. The insets in (c) and (g) enlarge the transport response around the Fermi energy. 
~\label{figs8}}
\end{figure}
To activate the anomalous transport response, we apply a 3\% uniaxial strain along the $a$ axis. This perturbation breaks the fourfold rotation and the diagonal symmetries connecting the two magnetic sublattices, removes the $[C_{2\perp}\Vert C_{4z}]$-enforced relation between the opposite-spin bands, and lifts their energetic equivalence, as evidenced by Fig.~\ref{figs8}(a). After SOC is included [Fig.~\ref{figs8}(b)], the Berry-curvature contributions from the two spin sectors no longer cancel, leading to finite energy-dependent anomalous Hall and anomalous Nernst conductivities [Figs.~~\ref{figs8}(c) and \ref{figs8}(d)]. Applying the same strain along the $b$ axis produces the symmetry-related band structure shown in Figs.~\ref{figs8}(e) and \ref{figs8}(f). Because the $a$- and $b$-strained configurations are transformed into each other by $C_{4z}\mathcal{T}$, their Berry curvatures have opposite signs, yielding $\sigma_{xy}^{(b)}(E)=-\sigma_{xy}^{(a)}(E)$ and $\alpha_{xy}^{(b)}(E)=-\alpha_{xy}^{(a)}(E)$, as demonstrated in Figs.~\ref{figs8}(g) and \ref{figs8}(h). Thus, simply rotating the strain direction by $90^\circ$ reverses both transverse responses, further confirming the general applicability of crystal-symmetry-controlled anomalous transport.

\color{black}

\newpage
\section{Magnetic configuration of materials}
In this section, we provide a detailed analysis of the magnetic configurations of the candidate materials and their evolution under strain.

\subsection{Magnetic ground state}
\begin{figure}[h]
\centering
\includegraphics[width=0.9\columnwidth]{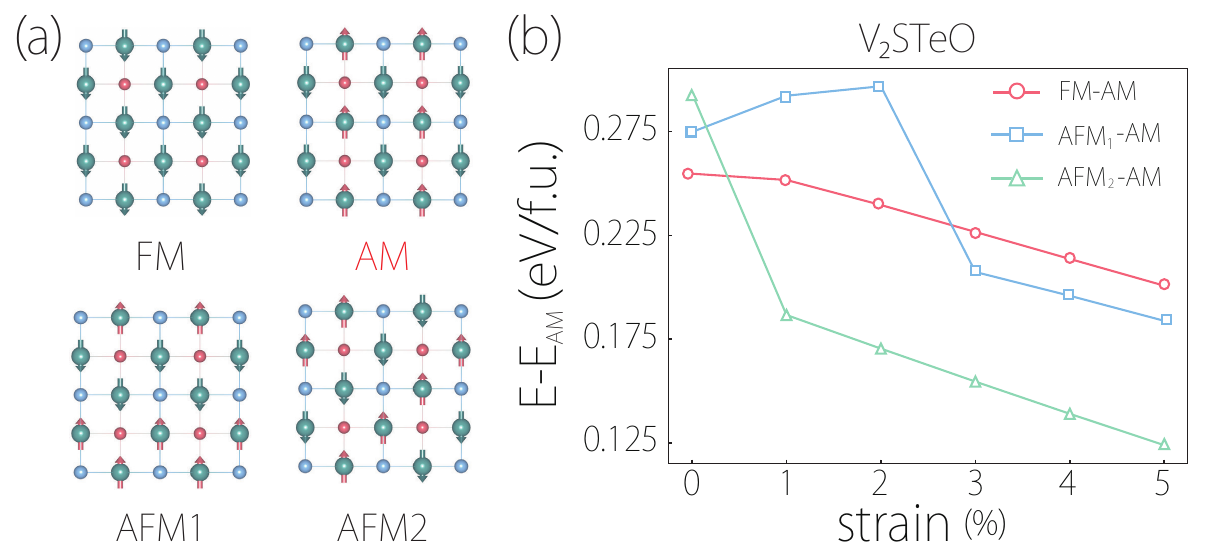} \caption{(a) Magnetic configurations of the ferromagnetic, altermagnetic, and two stripy antiferromagnetic (AFM1 and AFM2) orders of monolayer V$_{2}$STeO. (b) Energy differences of other magnetic configurations relative to the altermagnetic state under various uniaxial tensile strains.
~\label{figs9}}
\end{figure}

\begin{figure}[h]
\centering
\includegraphics[width=0.9\columnwidth]{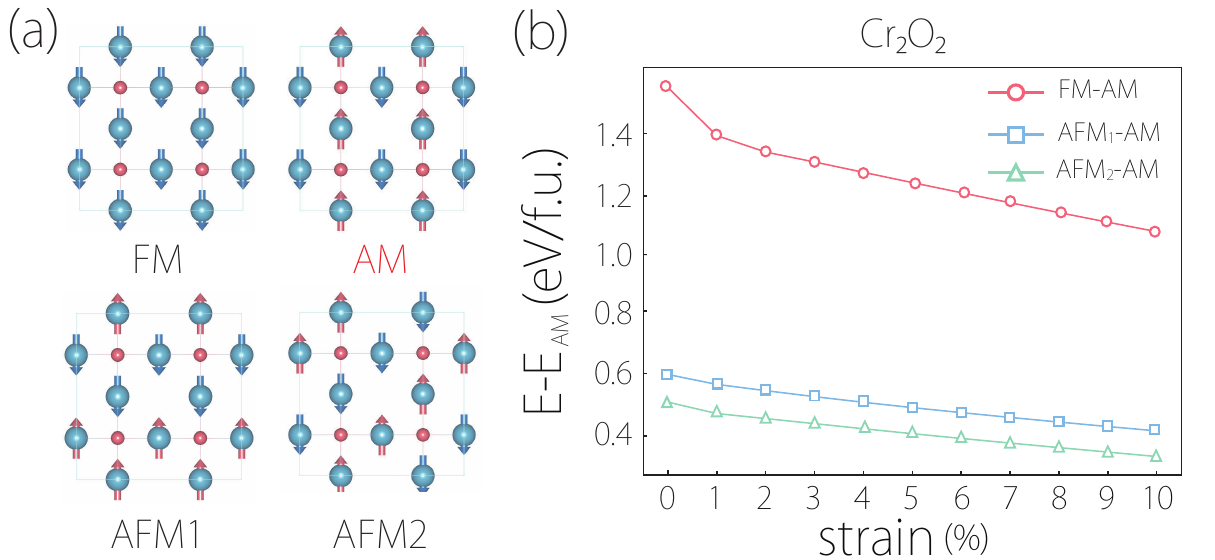} \caption{(a) Magnetic configurations of the ferromagnetic, altermagnetic, and two stripy antiferromagnetic (AFM1 and AFM2) orders of monolayer Cr$_{2}$O$_{2}$. (b) Energy differences of other magnetic configurations relative to the altermagnetic state under various uniaxial tensile strains.
~\label{figs10}}
\end{figure}

\begin{figure}[h]
\centering
\includegraphics[width=0.9\columnwidth]{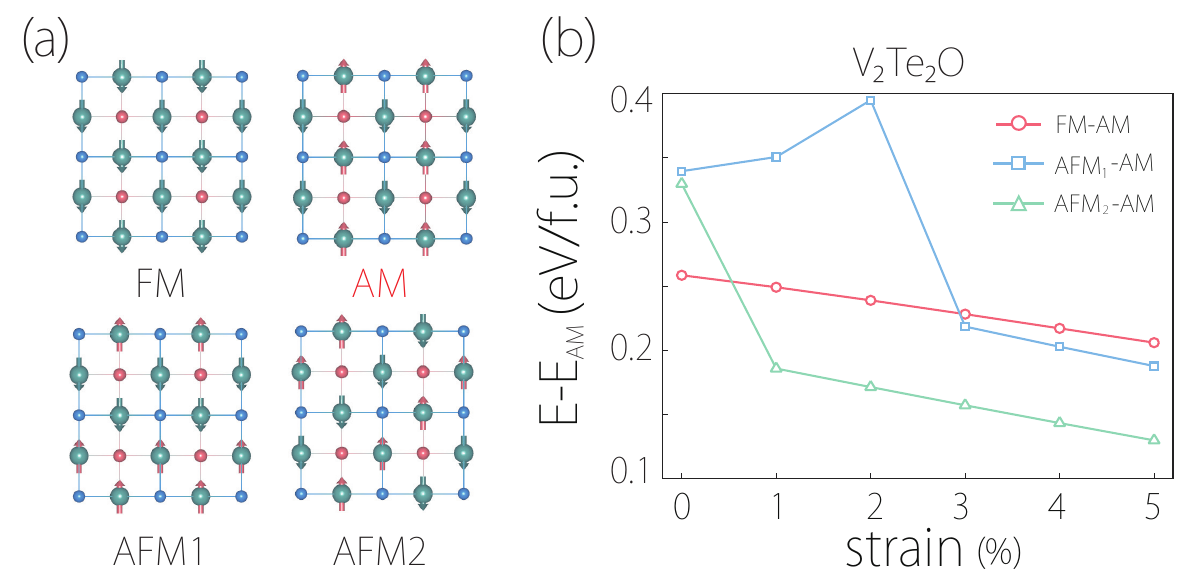} \caption{(a) Magnetic configurations of the ferromagnetic, altermagnetic, and two stripy antiferromagnetic (AFM1 and AFM2) orders of monolayer V$_{2}$Te$_{2}$O. (b) Energy differences of other magnetic configurations relative to the altermagnetic state under various uniaxial tensile strains.
~\label{figs11}}
\end{figure}

To identify the magnetic ground state of the V$_{2}$STeO, Cr$_{2}$O$_{2}$, and V$_{2}$Te$_{2}$O monolayers, we systematically compared the total energies of four representative magnetic configurations, namely the ferromagnetic (FM), altermagnetic (AM),  and two antiferromagnetic(AFM) states (denoted as AFM1 and AFM2), as illustrated in Figs.~\ref{figs9}(a), \ref{figs10}(a), and \ref{figs11}(a). Our calculations reveal that in the pristine state, these compounds energetically favor the AM state. Crucially, this AM configuration is highly stable under strain: V$_{2}$STeO, Cr$_{2}$O$_{2}$ and V$_{2}$Te$_{2}$O preserve their AM ground states under uniaxial tensile strains of at least up to 5\%, 10\%, and 5\%, respectively. The corresponding energy differences of other magnetic states relative to the AM state under various strains are presented in Figs.~\ref{figs9}(b), \ref{figs10}(b) and \ref{figs11}(b).

\subsection{Magnetocrystalline anisotropy energy}
\begin{figure}[h]
\centering
\includegraphics[width=0.8\columnwidth]{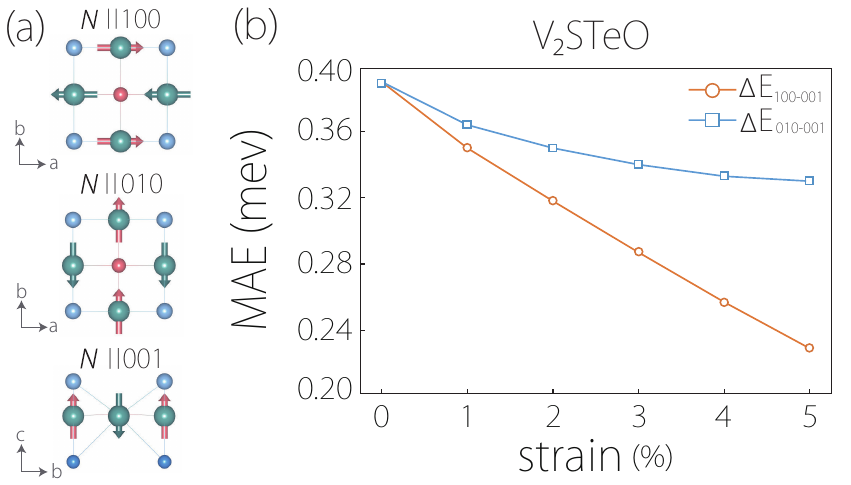} \caption{(a) Magnetic configurations of the altermagnetic order in V$_{2}$STeO with the Néel vector aligned along three high-symmetry directions. (b) Calculated magnetocrystalline anisotropy energy under various uniaxial tensile strains, defined as the energy difference between the in-plane and out-of-plane directions.
~\label{figs12}}
\end{figure}

\begin{figure}[h]
\centering
\includegraphics[width=0.8\columnwidth]{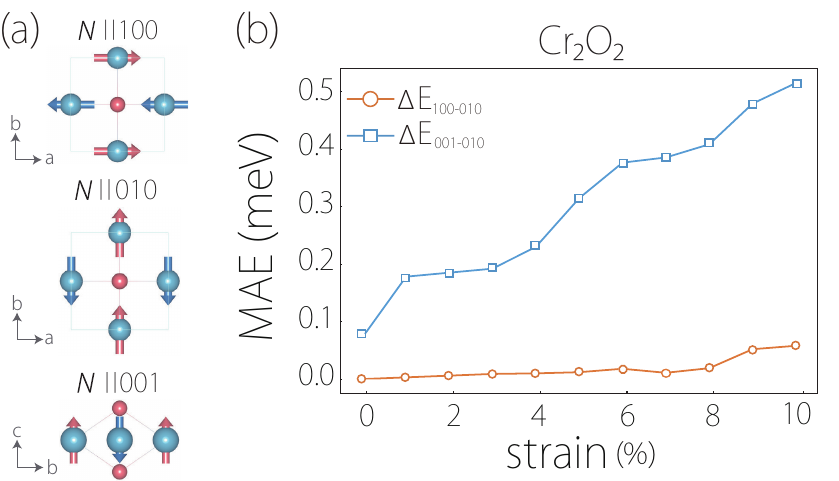} \caption{(a) Magnetic configurations of the altermagnetic order in Cr$_{2}$O$_{2}$ with the Néel vector aligned along three high-symmetry directions. (b) Calculated magnetocrystalline anisotropy energy under various uniaxial tensile strains, defined as the energy difference between the [001] ([100]) and [010] directions.
~\label{figs13}}
\end{figure}

\begin{figure}[h]
\centering
\includegraphics[width=0.75\columnwidth]{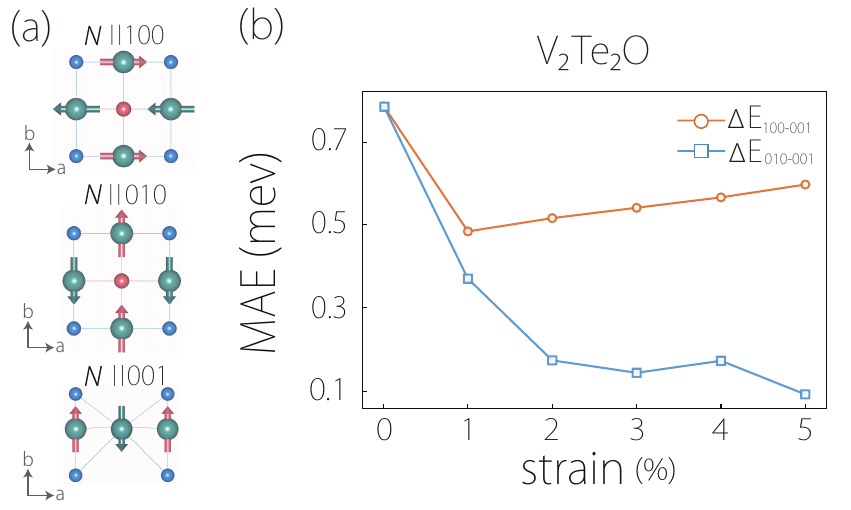} \caption{(a) Magnetic configurations of the altermagnetic order in V$_{2}$Te$_{2}$O with the Néel vector aligned along three high-symmetry directions. (b) Calculated magnetocrystalline anisotropy energy under various uniaxial tensile strains, defined as the energy difference between the in-plane and out-of-plane directions.
~\label{figs14}}
\end{figure}

In addition, we evaluate the magnetocrystalline anisotropy energy (MAE) of the AM ground states by comparing the total energies with the magnetization aligned along
three high-symmetry directions, namely, [100], [010], and [001], with SOC explicitly included [see Figs.~\ref{figs12}(a)-\ref{figs14}(a)]. Our calculations indicate that in the pristine state, V$_{2}$STeO and V$_{2}$Te$_{2}$O exhibit an out-of-plane easy axis along the [001] direction, and the out-of-plane magnetism in both materials remains robust under uniaxial tensile strain up to 5\% [see Fig.~\ref{figs12}(b) and \ref{figs14}(b)]. Such a stable out-of-plane Néel vector in altermagnets is crucial for realizing our proposed crystal-symmetry-controlled anomalous transport. For Cr$_{2}$O$_{2}$ [see Fig.~\ref{figs13}(a)], the system exhibits an in-plane magnetic easy axis, which is consistent with previous reports~\cite{chen2023giant}. In the pristine state, the energy difference between the in-plane and out-of-plane magnetic configurations is relatively small [see Fig.~\ref{figs13}(b)]. Therefore, we suggest applying an external magnetic field to align its Néel vector out of the plane~\cite{feng2022anomalous}.


\newpage

\newpage

\newpage
\color{black}
\bibliography{supref}